\documentclass[12pt,epsf]{article}
\usepackage{amssymb,amsmath}
\usepackage{graphicx}
\usepackage{color}
\usepackage[colorlinks=false,linkcolor=blue, citecolor=blue, linktoc=all]{hyper ref}

\parskip=\baselineskip

\newcommand{\be}{\begin{equation}}
\newcommand{\ee}{\end{equation}}
\newcommand{\bea}{\begin{eqnarray}}
\newcommand{\eea}{\end{eqnarray}}
\newcommand{\beas}{\begin{eqnarray*}}
\newcommand{\eeas}{\end{eqnarray*}}
\newcommand{\ba}{\begin{array}}
\newcommand{\ea}{\end{array}}

\newcommand{\cDsl}{{{\cal D}\kern-.65em /\,}}
\newcommand{\cDslsm}{{{\cal D}\kern-.5em /\,}}
\newcommand{\nabslsm}{\nabla\kern-.55em /}
\newcommand{\pasl}{\pa\kern-.55em /}
\newcommand{\psl}{p\kern-.45em /}
\newcommand{\Dsl}{D\kern-.65em /}
\newcommand{\Asl}{A\kern-.55em /}
\newcommand{\nabsl}{\nabla\kern-.65em /\kern+.2em}
\newcommand{\qsl}{q\kern-.5em /}
\newcommand{\ksl}{k\kern-.5em /}
\newcommand{\rsl}{r\kern-.5em /}

\newcommand{\cDslLCsq}{{\stackrel{\circ}{\cDsl^{\kern2pt 2}}}}

\newcommand\cc[1]{#1^{^{\kern-6pt \circ}}\kern2pt}

\newcommand{\pa}{\partial}

\newcommand{\beq}{\begin{equation}}
\newcommand{\eeq}{\end{equation}}
\newcommand{\beqn}{\begin{eqnarray}}
\newcommand{\eeqn}{\end{eqnarray}}

\def\dalemb#1#2{{\vbox{\hrule height .#2pt
\hbox{\vrule width.#2pt height#1pt \kern#1pt
\vrule width.#2pt}
\hrule height.#2pt}}}

\newcommand{\vx}{\vec{x}}

\newcommand{\vk}{\vec{k}}

\newcommand{\cO}{\mathcal{O}}

\begin{document}

\begin{titlepage}
\hfill
\vbox{
    \halign{#\hfil         \cr
           } 
      }  
\vspace*{0mm}
\begin{center}
{\Large \bf From Euclidean Sources to Lorentzian Spacetimes \\ in Holographic Conformal Field Theories}

\vspace*{10mm}
Don Marolf${}^a$, Onkar Parrikar${}^b$, Charles Rabideau${}^{b,c}$, \\ Ali Izadi Rad${}^d$, and Mark Van Raamsdonk$^d$
\vspace*{1cm}
\let\thefootnote\relax\footnote{marolf@physics.ucsb.edu,parrikar@sas.upenn.edu,rabideau@sas.upenn.edu,izadi@phas.ubc.ca,mav@phas.ubc.ca}

{${}^{a}$Department of Physics, University of California, \\
Santa Barbara, CA 93106, USA \\
${}^{b}$David Rittenhouse Laboratory, University of Pennsylvania,\\
209 S.33rd Street, Philadelphia PA, 19104, U.S.A. \\
${}^{c}$Theoretische Natuurkunde, Vrije Universiteit Brussel (VUB), and \\
International Solvay Institutes, Pleinlaan 2, B-1050 Brussels, Belgium${}^{d}$ \\
${}^{d}$Department of Physics and Astronomy,
University of British Columbia\\
6224 Agricultural Road,
Vancouver, B.C., V6T 1W9, Canada}

\end{center}
\begin{abstract}

We consider states of holographic conformal field theories constructed by adding sources for local operators in the Euclidean path integral, with the aim of investigating the extent to which arbitrary bulk coherent states can be represented by such Euclidean path-integrals in the CFT. We construct the associated dual Lorentzian spacetimes perturbatively in the sources. Extending earlier work, we provide explicit formulae for the Lorentzian fields to first order in the sources for general scalar field and metric perturbations in arbitrary dimensions. We check the results by holographically computing the Lorentzian one-point functions for the sourced operators and comparing with a direct CFT calculation. We present evidence that at the linearized level, arbitrary bulk initial data profiles can be generated by an appropriate choice of Euclidean sources. However, in order to produce initial data that is very localized, the amplitude must be taken small at the same time otherwise the required sources diverge, invalidating the perturbative approach.

\end{abstract}

\end{titlepage}

\tableofcontents

\section{Introduction}

According to the AdS/CFT correspondence \cite{maldacena1999large, Witten:1998qj, Gubser:1998bc}, certain conformal field theories provide a nonperturbative description of quantum gravitational theories describing physics in spacetimes that are asymptotically anti-de-Sitter. Different states of the conformal field theory correspond to different states of the gravitational theory, but only some of these states will have gravity dual descriptions well-represented by simple classical spacetimes. For example, other states might correspond to quantum superpositions of macroscopically different geometries. It is interesting to understand better the CFT characterization of the holographic/geometric states.

A related question is the following: given a spacetime geometry which is solution to the classical gravitational equations for some theory of gravity dual to a holographic CFT, can we explicitly describe the field theory state (or family of states) that are dual to this geometry, i.e. whose observables can be computed by gravity methods by perturbing about this geometry? Understanding these questions is important in order to improve our understanding of how gravitational physics emerges in holographic conformal field theories. For example, in trying to derive classical gravitational equations directly from the CFT, it should be important to work with states for which these equations are relevant.

In this paper, following various earlier works \cite{Skenderis:2008dh, Skenderis:2008dg, Botta-Cantcheff:2015sav, Christodoulou:2016nej}, we consider a large class of states in a holographic conformal field theory obtained by adding sources for local, primary operators (dual to classical bulk fields) to the Euclidean path integral defining the vacuum state of the CFT. Explicitly, we consider states with wavefunctionals defined as
\be
\label{PIstate}
\langle \phi_0 | \Psi_\lambda \rangle = \int^{\phi(\tau=0) = \phi_0}_{\tau<0} [d \phi(\tau)] e^{-S_E - \int_{-\infty}^0 d \tau \lambda_\alpha(x, \tau) {\cal O}_\alpha(x,\tau)} \; ,
\ee
where ${\cal O}_\alpha$ are operators dual to light fields in the bulk. In \cite{Skenderis:2008dh, Skenderis:2008dg}, it was noted that this construction defines states whose Lorentzian correlation functions can be computed via a dual gravity calculation. In \cite{Botta-Cantcheff:2015sav}, it was argued that these states define coherent states of the (nearly free) bulk fields, as we would expect for states describing classical field configurations. Recently, it was shown \cite{Faulkner:2017tkh} that for states of the form (\ref{PIstate}) in a large class of CFTs, there is always an associated classical geometry (defined to second order in perturbation theory) which captures the entanglement entropies of ball-shaped regions via the Ryu-Takayanagi formula \cite{Ryu:2006bv,Hubeny:2007xt} and which satisfies Einstein's equations perturbatively to second order. To obtain a classical bulk, here and below we take $\lambda$ to be parametrically of the same magnitude as the CFT action in the holographic large $N$ limt; e.g. $\lambda$ is of order the central charge $c$ for $d=2$ CFTs and of order $N^2$ for ${\cal N}=4$ super Yang-Mills in $d=4$.

In light of these results, it is clearly of interest to understand in more detail the map between Euclidean path-integral states (\ref{PIstate}) and Lorentzian spacetimes, provide additional checks that the states (\ref{PIstate}) are indeed holographic, and investigate to what extent an arbitrary classical bulk solution can be described by a state of the form (\ref{PIstate}). This will be the aim of the present paper.

We begin in section \ref{sec2} by reviewing the motivations for identifying these path integral states as holographic states. We start with  a heuristic argument, that states of this form correspond to coherent states of the bulk mode operators, as we would expect for states describing classical field configurations in the weakly coupled bulk theory. This was emphasized in \cite{Botta-Cantcheff:2015sav}. We then review the concrete prescription of Skenderis and van Rees \cite{Skenderis:2008dh, Skenderis:2008dg} for determining the Lorentzian geometry associated with a given state of the form (\ref{PIstate}). This involves finding a Euclidean solution associated with a CFT perturbed by sources $\lambda(x, \tau)$ as in (\ref{PIstate}) for $\tau < 0$ and $\lambda^*(x,-\tau)$ for $\tau > 0$, then slicing this solution open on a bulk spacelike slice that ends at $\tau=0$, reading off Lorentzian initial data from the slice, and finally finding a Lorentzian solution with this initial data.

Given this construction, we would like to understand better what class of spacetimes we can describe using states of this form. Can we associate a path-integral state to an arbitrary Lorentzian geometry? Are there multiple path-integral states which correspond to the same geometry? Are there some CFT states with a good classical gravity description that cannot be described or well-approximated in this way?

The prescription to associate a Lorentzian geometry with a given path integral state is generally a complicated nonlinear classical gravity problem. However, we can ask our questions already at a perturbative level where everything can be calculated explicitly. We perform this linearized study in section \ref{sec3}, generalizing work in \cite{Botta-Cantcheff:2015sav, Christodoulou:2016nej}. For a CFT on Minkowski space, states defined by (\ref{PIstate}) are expected to describe geometries which are perturbations of Poincar\'e-AdS spacetime.  We compute explicitly to first order in the sources the bulk initial data that defines the Lorentzian spacetime dual to the CFT state; we consider arbitrary sources for scalar operators or the stress-energy tensor in arbitrary dimensions. Using this solution and the usual holographic dictionary, we deduce the Lorentzian one-point function of CFT operators used in the definition of the state. We compare this with the one-point function computed from the sources by a direct CFT calculation and find complete agreement, providing a detailed check of the proposed mapping between states (\ref{PIstate}) and Lorentzian geometries.

Our results give an explicit identification between Fourier modes of the sources and modes of the bulk fields. Formally, we can invert the mapping to find (at the linearized level) the sources corresponding to an arbitrary Lorentzian solution (still working perturbatively). However, there is a subtlety: in section \ref{sec4}, we show that if we try to construct a sequence of bulk initial data functions that approaches a delta function, the sources diverge in the limit where the initial data becomes infinitely localized. Thus, from the bulk point of view, the validity of perturbation theory depends not only on the magnitude of the fields but also on how localized their features are. The conclusion is that while it is possible to generate any smooth field profile in the initial data, in order to ensure the validity of the perturbative approach, we need to make the amplitude smaller depending on how fine the features are.

An interesting result of our perturbative analysis is that the the one-point functions we calculate perturbatively are always functions whose Fourier modes obey $\omega \ge |k|$. In section \ref{sec5}, we show that this property holds non-perturbatively for the one-point function of any local scalar operator of fixed scaling dimension in any state of a CFT created by arbitrary insertions of scalar quasi-primary operators in the Euclidean path integral. Qualitatively this means that the time-variation of the one-point functions must be faster than the spatial variation. For example, while it is possible to have a one-point function which is localized in time but homogeneous in space, it is impossible for a one-point function to be localized in space but constant in time - the inhomogeneity necessarily spreads out. This property has of course been noted for holographic theories at leading order in $1/N$ before \cite{Bena:1999jv, Papadodimas:2012aq} - it is related to the condition that the bulk fields are constructed from modes that do not diverge as we move into the bulk.

We conclude in section \ref{sec6} with a discussion of possible future directions. Finally, in appendix \ref{appA} we comment on the perturbative map between CFT one-point functions and the sources required to produce them at higher orders in perturbation theory. This may be useful in extending our work beyond the linearized order.

\section{Bulk geometries from path integral states}\label{sec2}

We consider conformal field theories in $d$ spacetime dimensions. We assume that the field theories are holographic, and that the classical limit of the dual gravitational description is Einstein gravity coupled to scalar matter fields. Then for each matter field $\phi_\alpha$ in this gravitational theory, there is a low-dimension primary operator ${\cal O}_\alpha$ in the CFT such that the asymptotic behaviour of $\phi_\alpha$ in a given spacetime is related to the one-point function of ${\cal O}_\alpha$ in the dual CFT state.

Our focus will be the particular class of states (\ref{PIstate}); we will now review some general arguments that these states of the CFT should be dual to spacetimes with a good classical description.

\subsubsection*{Motivation: path integral states as bulk coherent states}

We start by considering the CFT on a spatial $S^{d-1}$ and recalling that the vacuum state can be constructed using a Euclidean path integral over $S^{d-1} \times \mathbb{R}^-$ as
\be
\langle \phi_0 | vac \rangle = \int^{\phi(\tau=0) = \phi_0}_{\tau<0} [d \phi(\tau)] e^{-S_E} \; .
\ee
where $\tau$ is the Euclidean time. By a conformal transformation, the half-cylinder $S^{d-1} \times \mathbb{R}^-$ can be mapped to a ball $B^d$ with the sphere at $\tau = - \infty$ mapping to the origin. According to the state-operator correspondence in CFTs, any state in the CFT Hilbert space on $S^{d-1}$ can be defined by a similar  Euclidean path integral on the Euclidean ball $B^d$ by inserting a local operator into the Euclidian path integral at the center of this ball (i.e. at  $\tau = - \infty$ in the original coordinates),
\be
\label{origin}
\langle \phi_0 | \Psi_{\cal O} \rangle = \int^{\phi(\partial B) = \phi_0}_{B^d} [d \phi] e^{-S_E} {\cal O}(0) \; .
\ee
Here, operators of fixed scaling dimension give rise to states with energy (measured in units of the inverse sphere radius) equal to this dimension. General states are produced by inserting linear combinations of these operators.

For a holographic CFT, consider the low-dimension primary operator ${\cal O}_\alpha$ associated with some bulk field $\phi_\alpha$. Inserting this operator at the origin into the path integral (\ref{origin}) produces a state corresponding to adding a single quantum to the lowest-energy mode of the field $\phi_\alpha$ in AdS \cite{Aharony:1999ti}. Exciting other modes of this field corresponds to inserting conformal descendants of ${\cal O}_\alpha$, which are derivative operators $\left[P_{i_1}, \left[P_{i_2},\cdots, \left[P_{i_n},\mathcal{O}_{\alpha}\right]\cdots\right]\right] = \partial_{i_1} \cdots \partial_{i_n} {\cal O}_\alpha(0)$. We can denote the corresponding bulk creation operators by $a^\dagger_{\alpha i_1 \cdots i_n}$. In radial quantization, the corresponding annihilation operators are given by $a_{\alpha; i_1\cdots i_n} \sim \left[K_{i_1}, \left[K_{i_2},\cdots, \left[K_{i_n},\mathcal{O}^{\dagger}_{\alpha}\right]\cdots\right]\right]$.

In order to describe classical bulk field configurations, weak coupling intuition would suggest that we want to consider coherent states of the various bulk modes. In a free field theory, the coherent state associated with commuting modes $a_\alpha^\dagger$ is
\be
e^{z_\alpha a_\alpha^\dagger - z_\alpha^* a_\alpha} | 0 \rangle = {\cal N}_z e^{z_\alpha a_\alpha^\dagger} |0 \rangle = {\cal N}_z \Big(|0 \rangle + z_\alpha a_\alpha^\dagger|0 \rangle +  {1 \over 2} z_{\alpha_1} z_{\alpha_1} a_{\alpha_1}^\dagger a_{\alpha_2}^\dagger|0 \rangle + \dots \Big)
\ee
where ${\cal N}_z$ is a normalization factor. Thus, the coherent state ``exponentiates'' the infinitesimal transformation that adds a general linear combination of modes $z_\alpha a_\alpha^\dagger|0 \rangle$ to the vacuum.\footnote{Mathematically, this is precisely the exponential map applied to the tangent vector on state space defined by $\delta |\psi \rangle = \sum_\alpha z_\alpha a_\alpha^\dagger|  vac \rangle$.} Since the bulk description for a large $N$, strongly coupled CFT is also weakly coupled, we expect that bulk classical field configurations can be described roughly as coherent states built from the bulk mode operators (we have left the sum over $\alpha, i_1, i_2 \cdots $ implicit below):
\be
{\cal N} e^{z_{\alpha i_1 \cdots i_n} a^\dagger_{\alpha i_1 \cdots i_n}}|0 \rangle
= {\cal N} \Big(|0 \rangle  +  z_{\alpha i_i \cdots i_n} a^\dagger_{\alpha i_1 \cdots i_n} |0 \rangle  + {1 \over 2} z_{\alpha i_1 \cdots i_n} z_{\alpha' i'_i \cdots i'_n} a^\dagger_{\alpha i_1 \cdots i_n}  a^\dagger_{\alpha' i'_1 \cdots i'_n} |0 \rangle \dots \Big)\; .
\label{coherent}
\ee

We will now see that precisely such an expression arises when we consider states
\be
\label{PIstateBall}
\langle \phi_0 | \Psi_\lambda \rangle = \int^{\phi(\partial B) = \phi_0}_{B^d} [d \phi] e^{-S_E - \int_{B^d} d^d x \lambda_\alpha(x) {\cal O}_\alpha(x)} \; ,
\ee
obtained by adding sources for the low-dimension primaries ${\cal O}_\alpha$ to the path integral.
In the path integral exponent, we can expand\footnote{Here and below, a sum over $n$ is implied.}
\bea
\int_{B^d} \lambda_\alpha(x) {\cal O}_\alpha(x) &=& \int_{B^d} d^d x \lambda_\alpha(x) {1 \over n!} x^{i_1} \cdots x^{i_n} \partial_{i_1} \cdots \partial_{i_n} {\cal O}_\alpha(0) \cr
&\equiv&  \lambda_\alpha^{i_1 \cdots i_n} \partial_{i_1} \cdots \partial_{i_n} {\cal O}_\alpha(0)
\eea
where $\lambda_\alpha^{i_1 \cdots i_n}$ are the multipole moments of the sources on the ball.

At the linear order in $\lambda$, we have that
\be
|\Psi_\lambda \rangle_{S^{d-1}} = \int_{B^d} [d \phi_\alpha] e^{-S_E}\left(1 - \lambda_\alpha^{i_1 \cdots i_n} \partial_{i_1} \cdots \partial_{i_n} {\cal O}_\alpha(0) + {\cal O}(\lambda^2) \right) \; .
\ee
As above, insertions of the conformal descendant operators $\partial_{i_1} \cdots \partial_{i_n} {\cal O}_\alpha$ add quanta of the various modes of the bulk field $\phi_\alpha$. Thus, we can roughly rewrite the state as\footnote{We could also split the operator $\partial_{i_1} \cdots \partial_{i_n} {\cal O}_\alpha$ into traceless and trace parts, and define creation operators corresponding to the various states in each of the irreducible representations of $SO(d)$.}
\be
\label{tangent}
|\Psi_\lambda \rangle_{S^{d-1}} = |0 \rangle -  \lambda_\alpha^{i_1 \cdots i_n} a^{\dagger}_{\alpha; i_1 \cdots i_n} |0 \rangle + {\cal O}(\lambda^2); .
\ee
which agrees with the leading order terms in (\ref{coherent}). Furthermore, the higher order terms in (\ref{coherent}) are defined by exponentiating the first order perturbation, and this is precisely what happens in the path integral expression (\ref{PIstateBall}). The higher order terms correspond to inserting multitrace operators at the origin, and these multitrace operators correspond to acting with multiple bulk creation operators as in the expansion of the coherent state.

Of course, in field theory, we have to be careful when dealing with operator products and introduce some type of regularization to define exactly what we mean by inserting multiple operators at a point. But this is precisely what our original expression (\ref{PIstateBall}) does for us: we can view it as a particular regularization of the naive expression that we would get by trying to translate bulk coherent states to a path integral expression using the operator/state correspondence. The discussion here is somewhat qualitative and meant primarily to provide intuition for why states (\ref{PIstateBall}) are related to classical spacetimes. One point that is not clear from this discussion is that the sources in (\ref{PIstateBall}) should vanish sufficiently rapidly at the edges of the ball in order to define a finite-energy state of the original theory. We will discuss this further below.

\subsubsection*{Semi-classical Schwinger-Keldysh}

We will now describe an explicit procedure, based on the work \cite{Skenderis:2008dg,Skenderis:2008dh} by Skenderis and van Rees, to determine the Lorentzian space time associated to a holographic CFT state of the form (\ref{PIstate}).

In the CFT, if we would like to calculate real time correlators
\be
\langle  \Psi_\lambda | {\cal O}_1(t_1) \cdots {\cal O}_n(t_n)|\Psi_\lambda \rangle = \langle  \Psi_\lambda | e^{- i H t_1} {\cal O}_1 e^{i H t_1} \cdots e^{-i H t_n} {\cal O} e^{i H t_n}|\Psi_\lambda \rangle
\ee
we can use the Schwinger-Keldysh formalism: we define a path integral that includes a Euclidean part (\ref{PIstate}) to define the state, the complex conjugate of this to define the ket, and Lorentzian parts corresponding to the real-time evolution operators that take us between the various times corresponding to the operator insertions. We include sources in the Lorentzian parts in order to define the generating functional for Lorentzian correlation functions. The correlations functions are calculated by taking derivatives with respect to these sources and then setting the sources to zero.

This path integral can be calculated holographically via a gravity problem in which we find a saddle-point geometry obtained by patching together Euclidean and Lorentzian parts corresponding to the various parts of the field theory contour. The geometry must satisfy Einstein's equations appropriate to the various parts, have asymptotic behaviour for the bulk fields consistent with the various sources, and satisfy certain matching conditions that ensure a smooth connection between the Euclidean and Lorentzian parts.

If we would only like to know what geometry is dual to the state (\ref{PIstate}), we can set the Lorentzian sources to zero. Then the Lorentzian geometry is obtained by time-evolving the initial data on the interface between the initial Euclidian part and the initial Lorentzian part with source-free boundary conditions. This initial data will be the same regardless of how large or small our Lorentzian contours are, so if our goal is just to determine the initial data, we can do away with the Lorentzian part altogether and consider the gravity dual of a Euclidian path integral with sources $\lambda_\alpha(\tau,x^i)$ for $\tau < 0 $ and $\lambda_\alpha^*(-\tau,x^i)$ for $\tau > 0$. The corresponding geometry is the solution to the bulk equations with asymptotic behavior corresponding to these sources. To define the Lorentzian initial data, we slice the geometry at $\tau = 0$ to read of $\phi_\alpha$ and $\partial_\tau \phi_\alpha$; this gives the Lorentzian initial data after adding an overall $i$ in the time derivative. For general asymptotically AdS spacetimes, we need to specify more precisely how this slicing and analytic continuation should be performed (going back to the detailed discussion in \cite{Skenderis:2008dg,Skenderis:2008dh}), since there are many ways to extend the boundary $\tau=0$ surface into the bulk. However, in this paper, we will work perturbatively in the sources, so that the cutting surface can be unambiguously chosen as the $\tau=0$ slice of the unperturbed bulk AdS spacetime.

In order to define the most general real Lorentzian spacetimes in this way, we will need to consider complex sources and thus complex Euclidean geometries. Again, this may lead to interesting conceptual issues when dealing with general states, but is straightforward so long as we are working perturbatively.

\section{Initial data and one-point functions to first order }\label{sec3}

In this section, starting from a general path-integral state with sources, we perform a holographic calculation as outlined in the previous section to determine explicitly the dual Lorentzian solution to first order in the sources. For simplicity, we first consider the case of scalar fields; analogous results for metric perturbations appear at the end of the section. Since we continue to treat the scalar fields as free and to ignore gravitational back-reaction, our results represent first-order perturbation theory in $\lambda$.  Recall in particular that we take $\lambda$ large at large $N$ so that large $N$ itself does not suffice to suppress non-linear corrections.

We start by considering the state (\ref{PIstate}) defined by adding a source $\lambda(\tau, \vec{x})$ for some scalar operator ${\cal O}$ of dimension $\Delta$.\footnote{Since we will be working at linear order in the sources, we can consider the sources for each field individually.} For a holographic CFT, this operator is associated with a bulk scalar field $\phi$ with mass $m^2 \ell^2 = \Delta (\Delta - d)$, where $\ell$ is the $AdS$ radius. To understand the Lorentzian spacetime dual to this state, our first step is to find the Euclidean asymptotically AdS spacetime with boundary conditions for fields determined by sources $\lambda$, extended to the full boundary via
\be
\lambda (\tau, \vec{x}) = \lambda^*(-\tau,\vec{x}) \; .
\ee
Using standard Poincar\'e coordinates
\be
ds^2 = {\ell^2 \over z^2} (dz^2 + dx_i dx_i + d \tau^2) \; ,
\ee
we have that the field $\phi$ satisfies
\bea
\label{scalarEOM}
\partial_\tau^2 \phi^E + \partial_i^2 \phi^E + \partial_z^2 \phi^E - (d-1) {1 \over z} \partial_z \phi^E - \left({m \ell \over z} \right)^2 \phi^E = 0 \; .
\eea
Using the standard Euclidean AdS/CFT dictionary \cite{Aharony:1999ti}, we have boundary conditions
\be
\label{Hol1}
\lim_{z \to 0} z^{d - \Delta} \phi^E(\tau,x,z) =  \lambda(\tau,x) \; .
\ee
We have used $\phi^E$ to indicate that we are considering a Euclidean solution. At first order in the sources, we can write the solution as
\beq
\phi^E(z,\tau,\vx) = \int_{-\infty}^{\infty} d\tau'\int d^{d-1} \vec{x}' \, K_{B\pa} ( z,\tau,\vec{x} | \tau',\vec{x}')\, \lambda(\tau',\vec{x}').
\eeq
where $K_{B\pa}$ is the bulk-to boundary propagator associated with the field $\phi^E$. The requirement that the solution not diverge in the interior fixes the propagator uniquely; it is given most easily as a Fourier decomposition:
\be
K_{B\pa} ( z,\tau,\vec{x} | \tau',\vec{x}') = \int {d^{d-1} k d \omega_E \over (2 \pi)^d} e^{-i \omega_E \tau' - i \vec{k} \cdot \vec{x}'} K(z,\tau,\vec{x}| \vec{k},\omega_E)
\ee
where
\be
K(z,\tau,\vec{x}|  \vec{k},\omega_E ) = { (k^2 + \omega_E^2)^{ \nu \over 2} \over 2^{\nu - 1} \Gamma(\nu)} z^{d \over 2} K_\nu(\sqrt{k^2 + \omega_E^2} \; z) e^{i \omega_E \tau + i\vec{k} \cdot \vec{x}} \; .
\ee
Here, $K_\nu$ is a Bessel function that decays exponentially for large $z$, and we have defined
\be
\nu = \sqrt{m^2 + {d^2 \over 4}} = \Delta - {d \over 2} \; .
\ee

\begin{figure}[t]
\centering
\includegraphics[height=6cm]{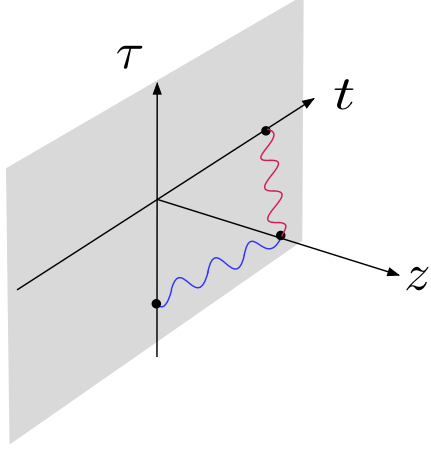}
\caption{\small{We prepare a state by turning on sources in Euclidean time $\tau$. This gives us some initial data on the $\tau=t=0$ surface in the bulk, which we can obtain from the bulk to boundary propagator (blue) and it's time derivative. Then, we can further evolve this data in real-time $t$ using the Lorentzian propagator (red) to obtain the real-time asymptotics, from which we read off the CFT one-point functions. Spatial directions in the CFT are supressed.} }
\end{figure}

To determine the Euclidean ``initial data'', we can then evaluate this bulk solution on the constant time slice at $\tau=0$:
\beq\label{ES1}
\phi^E(z,\tau=0,\vx) = \int_{-\infty}^{\infty} d\tau'\int d^{d-1} \vec{x}' \, K_{B\pa} ( z,0,\vec{x} | \tau',\vec{x}')\, \lambda(\tau',\vec{x}')
\eeq
\beq\label{ES2}
\pa_{\tau}\phi^E(z,\tau=0,\vx) = \int_{-\infty}^{\infty} d\tau'\int d^{d-1} \vec{x}' \, \pa_{\tau} K_{B\pa} ( z,0,\vec{x} | \tau',\vec{x}')\, \lambda(\tau',\vec{x}')
\eeq
Next, we analytically continue the initial data to real time
\be
\phi(z,x, t=0) = \phi^E(z,x, \tau = 0) \qquad \qquad \partial_t \phi(z,x,t=0) = i \partial_\tau \phi^E(z,x,\tau =0) \; ;
\ee
using the explicit form of the propagator, this gives
\bea
\phi(x,z,t=0) &=& { 1 \over 2^{\nu - 1} \Gamma(\nu)} \int {d^{d-1} \vec{k} d \omega_E \over (2 \pi)^d} \lambda_{\omega_E,\vk} z^{d \over 2} (k^2 + \omega_E^2)^{ \nu \over 2} K_\nu(\sqrt{k^2 + \omega_E^2} \; z) e^{i \vec{k} \cdot \vec{x}} \label{init} \\
\partial_t \phi(x,z,t=0) &=& - { 1 \over 2^{\nu - 1} \Gamma(\nu)} \int {d^{d-1} \vec{k} d \omega_E \over (2 \pi)^d} \omega_E \lambda_{\omega_E,\vk} z^{d \over 2} (k^2 + \omega_E^2)^{ \nu \over 2} K_\nu(\sqrt{k^2 + \omega_E^2} \; z) e^{i \vec{k} \cdot \vec{x}} \nonumber
\eea
as the explicit result for the Lorentzian initial data in terms of the Fourier modes of the sources. We wish to use these as initial data to evolve forward in real time using the Lorentzian bulk equations, setting the non-normalizable modes to zero at the AdS boundary (i.e. we assume there are no sources in real time).

We will proceed by writing a general solution to the linearized equations for the field $\phi$ and then matching to the initial data. It is convenient to express this in a basis of Fourier modes in the directions parallel to the boundary. Writing the basis functions as
\beq
\label{LorFour}
z^{d/2}\phi_{\omega,\vk}(z)e^{i\vk\cdot\vx- i\omega t} \; ,
\eeq
the Lorentzian version of the equations of motion (\ref{scalarEOM}) require that the functions $\phi_{\omega,\vk}$ satisfy a Bessel equation
\beq
z^2\pa_z^2\phi_{\omega,\vk}+z\pa_z\phi_{\omega,\vk}+z^2(\omega^2-\vk^2)\phi_{\omega,\vk}-\left(m^2+\frac{d^2}{4}\right)\phi_{\omega,\vk}=0 \; .
\eeq
Requiring that the solutions not diverge for large $z$ and also that they have the correct asymptotic behavior $z^{d/2} \phi_{\omega,\vk} \sim z^\Delta$ consistent with the absence of non-normalizable modes, we obtain solutions
\be
\label{defPhi}
\phi_{\omega,\vk}(z) =  \frac{2^{\nu}\Gamma(\nu+1)}{(\omega^2-\vk^2)^{\nu/2}} J_{\nu}(\sqrt{\omega^2-\vk^2}z)
\ee
with the restriction $\omega > |k|$. We have normalized by requiring that
\be
\label{PhiNorm}
\lim_{z\to 0}z^{-\Delta + {d \over 2}}\phi_{\omega,\vk} = 1 \; .
\ee
The Lorentzian solution arising from the initial data (\ref{init}) must be a linear combination of the modes (\ref{LorFour}), so we can write that
\be
\label{LorSol}
\phi_\lambda(z,x,t) = \int {d^{d-1} k d \omega \over (2 \pi)^d} C_{\omega,\vk} z^{d/2}\phi_{\omega,\vk}(z)e^{i\vk\cdot\vx- i\omega t} \; .
\ee
Now, using the initial data conditions and making use of the Bessel functions completeness relations
\beq
\label{Bessel1}
\int_0^{\infty} dz\, z\, J_{\nu}(kz) J_{\nu}(pz) = \frac{\delta (k-p)}{k} \; ,
\eeq
and the result
\beq
\label{Bessel2}
\int_0^{\infty} dz\,z\,J_{\nu}(kz) K_{\nu}(pz) =\frac{ \left(\frac{k}{p}\right)^{\nu}}{k^2+p^2} \; ,
\eeq
we find that
\be
\label{defC}
C_{\omega,\vk} =  {\rm sgn}(\omega) \theta(\omega^2-\vk^2)   { (\omega^2 - k^2)^\nu \over 2^{2 \nu} \Gamma(\nu) \Gamma(\nu + 1)} \int d \omega_E {\lambda_{\omega_E,\vk} \over \omega + i\omega_E}  \; .
\ee
where $\theta(x)$ is the usual step function vanishing for $x < 0$.

We can simplify the result further by writing it in terms of a Laplace transform for the time-dependence of the sources rather than a Fourier transform. Using the definition
\beq
\lambda_{\omega_E, \vk} = \int_{-\infty}^{\infty}d\tau\,e^{-i\omega_E \tau} \lambda_{\vk}(\tau) \; ,
\eeq
we can perform the $\omega_E$ integral in (\ref{defC}) by Cauchy's theorem to obtain
\be \label{R3}
C_{\omega,\vk}
= -  \;  2\pi \theta(\omega^2-\vk^2) \frac{(\omega^2-\vk^2)^{\nu}}{2^{2\nu}\Gamma(\nu)\Gamma(\nu+1)}\Big(\theta(\omega) {\cal L} \lambda_{\vec{k},\omega} +\theta(-\omega){\cal L} \lambda^*_{\vec{k},-\omega} \Big)
\ee
where we define the Laplace transform
\be
\label{defLT}
{\cal L} \lambda_{\vec{k},s} = \int_0^\infty d \tau \lambda_{\vec{k}}(-\tau) e^{-s \tau} \; .
\ee
The final result for the Lorentzian solution in terms of the sources is then given by (\ref{LorSol}), using (\ref{defPhi}) for the basis functions and either (\ref{defC}) or (\ref{R3}) for the coefficients.

Making use of our final expression for the Lorentzian solution, we can write a simpler result for the relation between sources and initial data. Defining the Bessel transform
\be
f_{\mu} = \int_0^\infty dz z J_\nu(\mu z) f(z)
\ee
we can use (\ref{R3}) in (\ref{LorSol}) to write
\bea
\phi_{\vec{k},\mu}(t=0) &=&  {\mu^\nu \over \Gamma(\nu) 2^\nu} {1 \over \sqrt{\vec{k}^2 + \mu^2}} \left({\cal L} \lambda_{\vec{k}, \sqrt{\vec{k}^2 + \mu^2}} + {\cal L} \lambda^*_{\vec{k}, \sqrt{\vec{k}^2 + \mu^2}} \right) \cr
 \partial_t \phi_{\vec{k},\mu}(t=0) &=&  i {\mu^\nu \over \Gamma(\nu) 2^\nu} \left({\cal L} \lambda_{\vec{k}, \sqrt{\vec{k}^2 + \mu^2}} - {\cal L} \lambda^*_{\vec{k}, \sqrt{\vec{k}^2 + \mu^2}} \right)
 \label{init1}
\eea
Thus, the Bessel transform of the spatial Fourier modes of the initial data functions are proportional to the Laplace transform of the spatial Fourier modes of the sources.

We can rearrange this to obtain
\be
{\cal L} \lambda_{\vec{k}, \sqrt{\vec{k}^2 + \mu^2}} = {\Gamma(\nu) 2^\nu \over \mu^\nu} \left(\sqrt{\vec{k}^2 + \mu^2}\phi_{\vec{k},\mu}(t=0) - i \partial_t \phi_{\vec{k},\mu}(t=0) \right) \; .
\ee
Using the definition (\ref{defLT}), we can formally invert the Laplace transform to obtain a formula for the sources in terms of the initial data:
\be
\label{LTinv}
\lambda_{\vec{k}}(\tau) = \int_{c- i\infty}^{c+i\infty} ds\; e^{-s\tau} {\Gamma(\nu) 2^\nu \over (s^2 - \vec{k}^2)^{\nu/2}} \left(s\phi_{\vec{k},\sqrt{s^2-\vec{k}^2}}(t=0) - i \partial_t \phi_{\vec{k},\sqrt{s^2-\vec{k}^2}}(t=0) \right).
\ee
where $c \in \mathbb{R}$ is to be chosen greater than the real part of the singularities of the integrand. When this gives a well-defined result, we can Fourier transform to deduce the position-space sources in terms of the initial data functions. Thus, for any initial data functions whose analyticity properties permit the inverse Laplace transform, we can find appropriate sources in the Euclidean path integral to produce this initial data and the corresponding Lorentzian solution. Unfortunately, it is not immediately clear which real initial data functions $\phi(\vec{x},z,t=0)$,$\partial_t \phi(\vec{x},z,t=0)$ we can obtain from functions with these analyticity properties. Thus, while we have a formal expression for the sources in terms of initial data, we will need additional analysis to understand whether essentially any Lorentzian solution at first order can be represented via Euclidian path-integral states as in (\ref{PIstate}). We will return to this analysis in section 4.

\subsubsection*{Holographic calculation of CFT one point functions in terms of sources.}

From our Lorentzian solution $\phi_\lambda$, we can read off the Lorentzian CFT one-point functions in the state (\ref{PIstate}) using the standard holographic dictionary.

The one-point function are determined holographically in terms of the normalizible modes of the asymptotic fields via \cite{Klebanov:1999tb}
\beq
\label{Hol2}
\lim_{z \to 0} z^{-\Delta}\phi (z,t,\vec{x}) = {1 \over 2 \Delta - d} \left\langle \cO(t,\vec{x})\right\rangle_{\psi} \; .
\eeq
so making use of (\ref{PhiNorm}) in (\ref{LorSol}), we find that the Fourier modes of the one-point functions are precisely
\bea
\langle {\cal O}_{\omega,\vec{k}} \rangle &=& (2 \Delta - d) \; C_{\omega, \vec{k}} \cr
&=&   {\rm sgn}(\omega) \theta(\omega^2-\vk^2)   {(2 \Delta - d)  (\omega^2 - k^2)^\nu \over 2^{2 \nu} \Gamma(\nu) \Gamma(\nu + 1)} \int d \omega_E {\lambda_{\omega_E,\vk} \over \omega + i\omega_E}
\label{defO}
\eea
where the coefficients $C_{\omega, \vec{k}}$ are given by (\ref{defC}) or (\ref{R3}).

This is the result to first order in $\lambda$ for large $N$ holographic CFTs. In part to check on the holographic prescription, we will compare it to a direct first-order calculation for general CFTs in the next section. In particular, we will reproduce the fact that all Fourier coefficients vanish for modes with $\omega < |\vec{k}|$ vanish.  In section 5, we will show that this continues to hold non-perturbatively, at least for a large class of states defined via the Euclidean path integral with arbitrary scalar operator insertions.

\subsection{Direct CFT calculation}

In the previous section, we have used the holographic Schwinger-Keldysh path integral to compute the relation between Euclidean sources and the dual Lorentzian solution to first order in the sources. From the solution, we used the holographic dictionary to read off the Lorentzian CFT one-point functions. As a check of the prescription and of our calculations, we will now show that the same results for one-point functions in terms of sources can be obtained through a direct CFT calculation. Since the CFT calculation will involve only two point functions, which have a universal form, the leading order result for one-point functions in terms of Euclidean sources is also universal for all CFTs.

\begin{figure}[t]
\centering
\includegraphics[width=\textwidth]{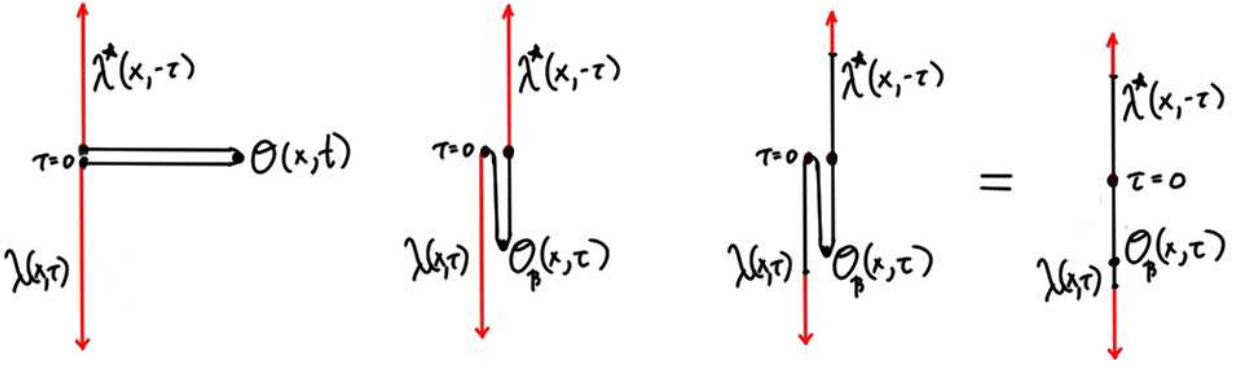}
\caption{\small{Path integral contours for evaluation of one point functions in Lorentzian time (first figure) and Euclidean time (second figure). Red contours indicate where sources have been turned on. To compute one-point functions at Euclidean times in a neighborhood of $\tau=0$ where the sources vanish, we can use a single Euclidean contour with $\tau \in (-\infty,\infty)$.}}
\label{PIfig}
\end{figure}

We start again with the path-integral state
\be
\label{PIstate2}
\langle \phi_0 | \Psi \rangle = \int^{\phi(\tau=0) = \phi_0}_{\tau<0} [d \phi(\tau)] e^{-S_E - \int_{-\infty}^0 d \tau \lambda_\alpha(x, \tau) {\cal O}_\alpha(x,\tau)} \; .
\ee
To evaluate the one point function for arbitrary Euclidean or Lorentzian time, we note that
\be
{\cal O}_\beta(x,\tau) = e^{ H \tau} {\cal O}_\beta(x,0) e^{-H \tau}
\ee
We would like to sandwich this operator between the bra and ket for the state (\ref{PIstate2}). The $e^{-H \tau}$ (or $e^{-i H t}$ for real times) gives an additional part of the path integral contour, as shown in figure (\ref{PIfig})
\be
\langle \Psi | {\cal O}_\beta(\hat{x},\hat{t}) | \Psi \rangle  = {1 \over Z_{\lambda}} \int_{C} [d \phi(\tau)] e^{\pm S_{E/L} - \int d^{d-1} x d \tau \lambda_\alpha (x, \tau) {\cal O}_\alpha(x,\tau)}{\cal O}_\beta(\hat{x},\hat{t})
\ee
where
\be
 Z_{\lambda} = \int [d \phi(\tau)] e^{-S_E - \int d^{d-1} x d \tau \lambda_\alpha(x, \tau)} \;
\ee
and the integral is defined on the appropriate contour $C$ as in figure (\ref{PIfig}) making the appropriate choices for $\pm$ and for Euclidean vs Lorentzian action. In this expression, we have extended the source to be defined for all real $\tau$ by $\lambda(\tau,x) =  \lambda^*(-\tau,x)$ for $\tau > 0$.

As shown in the third and fourth pictures in figure (\ref{PIfig}), if the sources are taken to vanish in some interval $[-\tau_0, \tau_0]$, we can evaluate the one point function for any operator in this interval simply by inserting it into the simple path integral without any additional parts to the contour. Assuming that the one point functions are analytic in the time, we can determine the one point functions for all Euclidean and Lorentzian times by analytically continuing from this region $[-\tau_0, \tau_0]$.

If the sources do not vanish on any interval around $\tau=0$, we can consider regulated sources, $\lambda_\epsilon$ defined to vanish in $[-\epsilon, \epsilon]$ and agree with $\lambda$ outside. In this case, we can analytically continue and then take the limit $\epsilon \to 0$. We will see below that this produces the correct results.

\subsubsection*{CFT result to first order in $\lambda$}

We will now use this general approach to write an expression for the one point function to first order in the sources. We will suppress reference to the spatial coordinates for now. First, we can rewrite the path integral expression as
\be
|\Psi_\lambda \rangle = {1 \over Z_\lambda^{1 \over 2}} {\large T} e^{-\int_{-\infty}^0 d \tau (H + \lambda(\tau) {\cal O})} |0 \rangle
\ee
Assuming the one point functions vanish in the vacuum state we can then expand this as
\be
|\Psi_\lambda \rangle = |0 \rangle - \int_{-\infty}^0 d \tau \lambda_\alpha (\tau) {\cal O}_\alpha (\tau) | 0 \rangle  + {\cal O} (\lambda^2)\; .
\ee
From this, we find to order $\lambda$
\bea
\langle \Psi_\lambda | {\cal O}_\beta(\hat{\tau}) | \Psi_\lambda \rangle &=& -\int_{-\infty}^0 d\tau \lambda_\alpha (\tau) \langle 0 | {\cal O}_\beta (\hat{\tau}) {\cal O}_\alpha(\tau) | 0 \rangle \\
&&  -\int_0^\infty d\tau \lambda_\alpha (\tau) \langle 0 |{\cal O}_\alpha (\tau) {\cal O}_\alpha(\hat{\tau})  | 0 \rangle \cr
&=&
-\int_{- \infty}^\infty d\tau \lambda_\alpha (\tau) \langle 0 | {\large} T \left\{ {\cal O}_\alpha (\tau) {\cal O}_\beta(\hat{\tau})  \right\} | 0 \rangle -\int_0^{\hat{\tau}} d\tau \lambda_\alpha (\tau) \langle 0 | [ {\cal O}_\alpha (\tau), {\cal O}_\beta(\hat{\tau})] | 0 \rangle \nonumber
\eea
where in the last line, the first term involves the standard two-point function computed via the Euclidean path integral, while the second term (a consequence of the extra parts of the contour in figure \ref{PIfig}) takes into account the non-time ordered parts.

This second term is absent if the operator lies in a neighborhood of $\tau=0$ where the sources vanish. In this case, for a CFT in $d$ dimensions, if ${\cal O}_\beta$ is dimension $\Delta$ and we have chosen a diagonal basis of operators
\be
\langle {\cal O}_\alpha(x_1) {\cal O}_\beta(x_2) \rangle =  {C_\alpha \delta_{\alpha \beta} \over |x_1 - x_2|^{2 \Delta_\alpha}} \; ,
\ee
we get
\be
\langle \Psi_\lambda | {\cal O}_\beta(\hat{\tau},\hat{x}) | \Psi_\lambda \rangle = -C_\beta \int d \tau d^{d-1} x \lambda_\beta (\tau,x) {1 \over ((\tau - \hat{\tau})^2 + (x - \hat{x})^2)^\Delta}
\ee
Here, $C_\beta$ is a normalization constant that we will specify below. Taking $\hat{\tau} = 0$, we see that the result for the one-point function diverges unless the sources vanish at $\tau=0$ unless $\Delta < {d \over 2}$. For an operator of dimension $\Delta \ge {d \over 2}$, finite one-point functions require that $\lambda$ vanishes at $\tau = 0$ as $\tau^n$ for $n > 2 \Delta - d$.

Starting from this formula, we can analytically continue to find an expression for the Lorentzian one-point functions. We get
\be
\label{Genlor}
\langle \Psi_\lambda | {\cal O}_\beta(t,\hat{x}) | \Psi_\lambda \rangle = - C_\beta \int d \tau d^{d-1} x \lambda_\beta (\tau,x) {1 \over ((\tau - it)^2 + (x - \hat{x})^2)^\Delta} \; .
\ee
We will argue below that this should be valid as long as the sources vanish sufficiently rapidly at $\tau=0$ so that the one-point functions are finite.

\subsubsection*{One-point functions from sources in Fourier space}

In order to compare with our holographic results, we would like to rewrite the general result (\ref{Genlor}) in terms of Fourier modes to check whether it matches (\ref{defO}). Defining
\be
{\cal O}(\vec{k}, \omega) = \int d^{d-1}x d t e^{-i \vec{k} \cdot \vec{x}} e^{-i \omega t} \langle \Psi_\lambda | {\cal O}_\beta(t,\hat{x}) | \Psi_\lambda \rangle \; ,
\ee
and
\be
\lambda(\vec{k}, \omega) = \int d^{d-1}x d t e^{-i \vec{k} \cdot \vec{x}} e^{-i \omega \tau} \lambda(\tau,\vec{x}) \; ,
\ee
we find that
\be
\label{OlambdaK}
{\cal O}(\vec{k}, \omega) = {1 \over 2 \pi} \int d \tau \int d \hat{\omega} \lambda(\vec{k},\hat{\omega}) e^{i \hat{\omega} \tau} K_\Delta (\vec{k},\omega,\tau)
\ee
where
\be
\label{defKdelta}
K_\Delta (\vec{k},\omega,\tau) = C_\beta \int d^{d-1} \vec{x} dt e^{-i \vec{k} \cdot \vec{x}} e^{-i \omega t} {1 \over (\vec{x}^2+(\tau - it)^2)^\Delta} \; .
\ee
To evaluate $K$, we define $x$ to be the spatial coordinate in the direction of $\vec{k}$ and integrate over the remaining spatial directions to obtain
\be
K_\Delta (\vec{k},\omega,\tau) = {C_\beta \pi^{{d \over 2}-1} \Gamma[\Delta - {d \over 2} + 1] \over \Gamma[\Delta]} \int dt e^{-i \omega t} \int d x  {e^{-i |\vec{k}| x}  \over (x^2+(\tau - it)^2)^{\Delta- {d \over 2} + 1}} \; .
\ee
The $x$ and $t$ integrals can be performed via contour integration. We will work with integer $\nu \equiv \Delta - {d \over 2}$ and analytically continue at the end. To compute the $x$ integral we close the contour in the lower-half plane and use
\be
\int_{-\infty}^\infty  {e^{-i |\vec{k}| x}  \over  (x^2+(\tau - it)^2)^{\nu + 1}} =  - 2 \pi i {1 \over \Gamma[\nu + 1]} \left\{ \ba{ll} \left. {d^\nu \over dx^\nu}  {e^{-i |\vec{k}| x}  \over  (x -i\tau - t)^{\nu + 1}} \right\vert_{x = -i \tau - t} & \tau > 0 \cr
\left. {d^\nu \over dx^\nu}  {e^{-i |\vec{k}| x}  \over  (x +i\tau +t)^{\nu + 1}} \right\vert_{x = i \tau + t} & \tau < 0 \ea \right.
\ee
The derivatives here give rise to $\nu + 1$ terms. For $\tau >0$, each of these gives a $t$ integral of the form
\be
\int_{-\infty}^\infty e^{- i \omega t} {e^{-i |k| (-i \tau - t)} \over (-t - i \tau)^{\nu + 1 + l}} = (2 \pi i) \theta(\omega - |k|) e^{- \omega \tau} (i (\omega - |k|))^{\nu + l} {1 \over (\nu + l)!}
\ee
where we again use contour integration. The result for $\tau < 0$ is
\be
\int_{-\infty}^\infty e^{- i \omega t} {e^{-i |k| (i \tau + t)} \over (t + i \tau)^{\nu + 1 + l}} = (2 \pi i) \theta(-\omega - |k|) e^{- \omega \tau} (-i (\omega + |k|))^{\nu + l} {1 \over (\nu + l)!} \; .
\ee
With these results, the sum of $\nu + 1$ terms is simply a binomial sum that is easily evaluated. The results for either sign of $\tau$ can be written as
\be
\label{Kdelta}
K_\Delta (\vec{k},\omega,\tau) = {C_\beta \pi^{{d \over 2}+1} \over 2^{2 \Delta - d-1} \Gamma[\Delta] \Gamma[\Delta - {d \over 2} +1]} e^{- \tau \omega} \theta({\rm sgn} (\tau) \omega - |k|)(\omega^2 - |k|^2)^{\Delta - {d \over 2}}
\ee
Inserting this into the result (\ref{OlambdaK}) and performing the integral over $\tau$, we get
\be
{\cal O}(\vec{k}, \omega) = C_\beta {\pi^{d \over 2} {\rm sgn}(\omega) \theta(\omega^2 - |k|^2) (\omega^2 - |k|^2)^{\Delta - {d \over 2}} \over 2^{2 \Delta - d} \Gamma[\Delta] \Gamma[\Delta - {d \over 2} +1]} \int d \hat{\omega} \lambda(\vec{k},\hat{\omega}) {1 \over \omega - i \hat{\omega}} \;
\ee
This matches precisely with the holographic result (\ref{defO}) as long as
\be
C_\beta = {\Gamma(\Delta) (2 \Delta - d) \over \pi^{d \over 2} \Gamma(\Delta - {d \over 2})} \; .
\ee
We can check (see for example \cite{Klebanov:1999tb}) that this is indeed the normalization of the two-point function consistent with the standard holographic dictionary (\ref{Hol1}) and (\ref{Hol2}). Thus, we have a detailed check of the prescription for defining holographic excited states, extending the work in \cite{Botta-Cantcheff:2015sav,Christodoulou:2016nej}.

\subsection{Results for metric perturbations}

We now consider the case of sources for the stress tensor i.e. metric perturbations on the Euclidean space where our path integral is defined. We denote the sources as ${1 \over 2} \gamma_{\mu \nu}(\tau, \vec{x}) T^{\mu \nu}$. Here, we have a redundancy in the description, since some metric perturbations are pure-gauge. Also, since we are dealing with a conformal field theory, two metric perturbations that describe conformally related boundary geometries should yield the same CFT state. Equivalently, we can say that since the stress-energy tensor is a conserved and traceless operator, sources for $T^\mu {}_\mu$ or $\partial_\mu T^{\mu \nu}$ should have no effect perturbatively. We could choose to restrict the form of $\gamma$ in order to avoid these redundancies; however, we will leave the form of $\gamma$ general so that our results are applicable to any chosen description of the boundary geometry.

As before, our first step is to find the Euclidean asymptotically AdS spacetime with boundary conditions for fields determined by sources $\gamma$, extended to the full boundary via
\be
 \gamma_{\mu \nu} (\tau, \vec{x}) =  \gamma_{\mu \nu}^*(-\tau,\vec{x}) \; .
\ee
We will describe the bulk metric perturbation using Fefferman-Graham coordinates,
\be
ds^2 = {\ell^2 \over z^2} (dz^2 + dx_i dx_i + d \tau^2 + H_{\mu \nu}(z,x,\tau) dx^\mu dx^\nu) \; .
\ee
For this choice, the bulk equations give
\bea
\partial_z \left( { 1 \over z} \partial_z H^\mu {}_\mu \right) &=& 0 \label{HE1}\\
\partial_z \left( \partial_\mu H^\mu {}_\nu - \partial_\nu H^\alpha {}_\alpha \right) &=& 0 \label{HE2}\\
z^{d-1} \partial_z \left( {1 \over z^{d-1}} \partial_z H_{\mu \nu} \right)  + \partial^2 H_{\mu \nu} &=& \eta_{\mu \nu} \partial_z H^\alpha {}_\alpha + \partial_\nu \partial_\alpha H^\alpha {}_\nu + \partial_\nu \partial_\alpha H^\alpha {}_\mu - \partial_\mu \partial_\nu H^\alpha {}_\alpha \label{HE3}
\eea
We need to solve these with boundary conditions
\be
\label{BCgamma}
\lim_{z \to 0}  H_{\mu \nu}(z,x,\tau) =  \gamma_{\mu \nu}(\tau, \vec{x}) \; .
\ee
From (\ref{HE1}) and (\ref{HE2}), we have that
\be
H_\mu {}^\mu = A + B z^2 \qquad \qquad \partial_\mu H^\mu {}_\nu = C_\nu + \partial_\nu A + \partial_\nu B z^2
\ee
where $A$, $B$, and $C$ depend on $x$ and $\tau$. Here, $A$ and $C$ are determined by our boundary conditions, and $B$ is determined from $C$ by the trace of equation (\ref{HE3}). We end up with
\be
A = \gamma^\mu {}_\mu \qquad C_\nu = \partial^\mu \gamma_{\mu \nu} - \partial_\nu \gamma_\alpha {}^\alpha \qquad B = {1 \over 2 (d-1)} \partial^2 \gamma_\alpha {}^\alpha -  {1 \over 2 (d-1)} \partial^\alpha \partial^\beta \gamma_{\alpha \beta}
\ee
Using these, the terms on the right side of equation (\ref{HE3}) can be expressed directly in terms of the sources $\gamma_{\mu \nu}$. Now, working in momentum space, we can solve (\ref{HE3}) with the boundary condition (\ref{BCgamma}) and the requirement that the solution should not blow up in the interior. Making use of Lorentz invariance to write the possible tensor structures, we find in the end that\footnote{This expression can be simplified considerably by using the redundancy in the sources described at the beginning of this section to put $\gamma_{\mu \nu}$ in a particular gauge; for example, choosing $\gamma$ to traceless and divergenceless eliminates all but the first term.}
\bea
H_{\mu \nu}(k^\mu,z) &=& \gamma_{\mu \nu}(k) \Big[ F(k z) \Big] \cr
&& + k^\alpha k_{(\mu} \gamma_{\nu) \alpha}(k) \Big[2 - 2 F(k z) \Big] \cr
&& + {k_\mu k_\nu \over k^2} \gamma_\alpha {}^\alpha(k) \Big[-{1 \over d-1} - {1 \over 2(d-1)} k^2 z^2 + {1 \over d-1} F(k z) \Big] \cr
&& + \eta_{\mu \nu} \gamma_\alpha {}^\alpha(k) \Big[{1 \over d-1} - {1 \over d-1} F(k z) \Big] \cr
&& + {k_\mu k_\nu k^\alpha k^\beta \over k^4} \gamma_{\alpha \beta}(k) \Big[-{d-2 \over d-1} + {1 \over 2(d-1)} k^2 z^2 + {d-2 \over d-1} F(k z) \Big] \cr
&& + \eta_{\mu \nu} { k^\alpha k^\beta \over k^2} \gamma_{\alpha \beta}(k) \Big[-{1 \over d-1} + {1 \over d-1} F(k z) \Big]
\eea
where $k^\mu = (\omega_E, \vec{k})$, $k = \sqrt{\omega_E^2 + \vec{k}^2}$ and $F$ is defined in terms of a Bessel function as
\be
F(x) = {x^{d \over 2} \over 2^{{d \over 2} - 1} \Gamma \left({d \over 2}\right)} K_{d \over 2}(x) \; .
\ee
This is normalized so that $F(x) \to 1$ as $x \to 0$.

The Lorentzian initial data is obtained by analytic continuation from the Euclidean metric perturbation at $\tau=0$:
\bea
H_{tt}(k,t=0,z) &=& - \int {d \omega_E \over 2 \pi} H_{\tau \tau} (k, \omega_E, z) \qquad \partial_t H_{tt}(k,t=0,z) = -i \int {d \omega_E \over 2 \pi} \partial_\tau H_{\tau \tau} (k, \omega_E, z) \cr
H_{ti}(k,t=0,z) &=& i \int {d \omega_E \over 2 \pi} H_{\tau i} (k, \omega_E, z) \qquad \partial_t H_{ti}(k,t=0,z) = - \int {d \omega_E \over 2 \pi} \partial_\tau H_{\tau i} (k, \omega_E, z) \cr
H_{ij}(k,t=0,z) &=& \int {d \omega_E \over 2 \pi} H_{ij} (k, \omega_E, z) \qquad \partial_t H_{ij}(k,t=0,z) = i \int {d \omega_E \over 2 \pi} \partial_\tau H_{ij} (k, \omega_E, z)
\label{initT}
\eea
We can now write a general solution to the Lorentzian equations for the metric perturbation and then match to this initial data. Requiring that the solutions not diverge for large $z$ and also that they have the correct asymptotic behavior $H \sim z^{d}$ consistent with the absence of non-normalizable modes (i.e. Lorentzian sources), we find
\be
H_{\mu \nu}(z,\vec{k},t) = \int {d \omega \over 2 \pi} e^{-i \omega t} {2^{d \over 2} \Gamma \left(1 + {d \over 2} \right) \over (\omega^2 - \vec{k}^2)^{d \over 4}} z^{d \over 2} J_{d \over 2}(\sqrt{ \omega^2 - \vec{k}^2} z) t_{\mu \nu}(\vec{k}, \omega) \; .
\ee
Here, the coefficients $t_{\mu \nu}$ are determined in terms of the sources $\gamma_{\mu \nu}$ by the matching conditions (\ref{initT}). We can extract them using the Bessel function integrals (\ref{Bessel1}) and (\ref{Bessel2}) above, and the additional relations
\beq
\int_0^{\infty} dz\,z^{1 - {d \over 2}}\,J_{d \over 2}(z)  =  {1 \over 2^{{d \over 2} - 1}} \Gamma \left({d \over 2} \right) \qquad \qquad \int_0^{\infty} dz\,z^{3 - {d \over 2}}\,J_{d \over 2}(z)  = {1 \over 2^{{d \over 2} - 3}} \Gamma \left({d \over 2}  -  1 \right) \; .
\eeq
Following analogous steps as in the scalar case, we could obtain expressions analogous to (\ref{defC}) and (\ref{R3}) expressing $t_{\mu \nu}$ directly as an integral transform of the sources; this is most straightforward in a gauge where $\gamma_{\mu \nu}$ is taken traceless and divergenceless.

Finally, the one-point function of the stress tensor in the CFT can be read off from the Lorentzian solution as
\beq
\label{HolT}
\langle T_{\mu \nu} \rangle = {d \ell^{d-3} \over 16 \pi G_N} t_{\mu \nu} \; .
\eeq

\section{Can we obtain arbitrary initial data?}\label{sec4}

In the previous section, we worked out the general relation (\ref{init1}) between path-integral sources and bulk initial data, working at linear order in the sources. We found a simple algebraic relation between the mixed Laplace/Fourier transform of the sources and a Bessel/Fourier transform of the initial data. We formally inverted this relation to express the sources directly in terms of the initial data as (\ref{LTinv}). Thus, naively, for any Lorentzian solution we wish to generate, we can use this result to find some corresponding initial data. However, the existence of the inverse Laplace transform appearing in (\ref{LTinv}) requires that we have an analytic continuation of the transformed initial data functions and also that this analytic continuation obeys certain constraints. It is not immediately clear what the implications of these constraints are for the (real) initial data functions.

A separate concern is whether the sources defined by the inversion formula (\ref{LTinv}) are sufficiently well behaved to justify our use of perturbation theory. For example, if for some choice of initial data, the sources defined by (\ref{LTinv}) diverge (e.g. for $\tau \to -\infty$), then the restriction to the linearized bulk equations that we have used is likely not justified. In this case, understanding whether the given initial data can be obtained from a path-integral state may require studying the full non-linear gravity equations in the bulk.

In this section, we will try to gain insight on these questions by asking to what extent it is possible to choose sources in order to approximate a delta function in the initial data. If we succeed in finding sources that lead to delta function initial data, we can then take linear combinations of such sources for various locations of the delta function to approximate any function in the initial data.

\subsection{Sources for localized initial data}

For our detailed analysis, we will specialize to the particularly simple case where the source is for a scalar operator with dimensions $\Delta = (d+1)/2$. Here, after a field redefinition
\be
\phi(z,x,t) = \frac{z^{d-1 \over 2}}{2\pi} \chi(z,x,t)
\ee
the field $\chi$ satisfies the ordinary Minkowski-space wave equation
\be
\partial_\mu \partial^\mu \chi  + \partial_z^2 \chi = 0 \; ,
\ee
and so many of the results simplify. In particular, the expression (\ref{init}) for the initial data in terms of the sources can be expressed as
\bea
\label{chifour}
\chi_{\vec{k}}(z,t=0) &=& \int_0^\infty {d \omega }\;  \lambda_{\omega, \vec{k}} \; e^{-\sqrt{\vec{k}^2 + \omega^2}z} \cr
\partial_t \chi_{\vec{k}}(z,t=0) &=& -\int_0^\infty {d \omega } \; \omega\; \lambda_{\omega, \vec{k}} \; e^{-\sqrt{\vec{k}^2 + \omega^2}z}
\eea
where $\lambda_{\omega, \vec{k}}$ is the Fourier transform of the sources and $\chi_{\vec{k}}(z)$ represents the Fourier transform of the initial data along the spatial field theory directions.

\subsubsection{Localizing initial data in the radial direction}

Let's consider first the case of sources and initial data that are translation-invariant in the field theory directions (i.e. have $\vec{k}=0$). In this case,
\bea
\label{init3}
\chi_{0}(z,t=0) &=& \int_0^\infty {d \omega}\;  \lambda_{\omega, 0} \; e^{-\omega z} \cr
\partial_t \chi_{0}(z,t=0) &=& -\int_0^\infty {d \omega} \; \omega\; \lambda_{\omega, 0} \; e^{-\omega z} \; .
\eea
If the initial data functions admit an analytic continuation to Euclidean time, we can check that these expressions are equivalent to
\be
\label{data_to_sources}
\lambda(\tau) = {1 \over 4 \pi} (\chi(z=i \tau) + \chi(z=-i \tau)) - {i \over 4\pi} \int_0^\tau( \partial_t \chi(z=i \tau) + \partial_t \chi (z=-i \tau)) d \tau
\ee
Focusing on the case of time-symmetric initial data (corresponding to real sources), we have simply \be
\label{data_to_sources1}
\lambda(\tau) = {1 \over 4\pi} (\chi(z=i \tau) + \chi(z=-i \tau)) \; ,
\ee
so the sources can be read off directly from an analytic continuation of the initial data.

Since we can approximate any initial data function $\chi(z)$ arbitrarily well by analytic functions (e.g. by taking a linear combination of Gaussians that approximate delta functions), the formula (\ref{data_to_sources1}) gives sources for essentially arbitrary initial data. However, as we anticipated in the introduction to this section, the sources obtained in this way can be poorly behaved. For example, if the initial data includes a Gaussian, the required sources diverge as $e^{c\tau^2}$ for $\tau \to - \infty$. This most likely invalidates our use of perturbation theory.

A more useful question is how well we can approximate arbitrary initial data using sources that obey some type of boundedness condition. To investigate this, we will ask how well we can approximate a delta function in the $z$ direction using sources whose $L_2$ norm is fixed to some specific value.\footnote{We could have chosen some other boundedness condition, choosing a different norm or requiring that the maximum value of the source never exceeds some value. We focus on the $L_2$ norm for convenience.
We should note the natural appearance of the Hardy norm: $\mathbb{H}_n (f) \equiv (\int_C |f(z)|^n dz)^{\frac{1}{n}}$ for a vertical contour $C$ in the complex plane parallel to the imaginary axis. The $\mathbb{H}_2$ norm of the produced initial data function is related to the $L_2$ norm of the source by a standard result in the theory of Laplace transforms,
$$
\int_{-\infty}^\infty dz |\chi(iz)|^2 = \int_{-\infty}^\infty dz d\omega d\omega' e^{-i(\omega-\omega')z} \lambda^*(\omega')\lambda(\omega) =2\pi \int_{-\infty}^\infty |\lambda(\omega)|^2 \,.
$$
However, the Hardy norm seems unrelated to the validity of bulk perturbation theory, so we will not consider it any further. }

\subsubsection{Variational calculation}

To address this question, we consider the variational problem of minimizing the variance
\be
V = {\int_0^\infty dz |\chi(z)|^2 (z - z_0)^2 \over \int_0^\infty dz |\chi(z)|^2}
\ee
subject to
\be
\label{norm2}
\frac{\int d \omega |\lambda(\omega)|^2 }{\int_0^\infty dz |\chi(z)|^2}  = 2 \pi \frac{\int d \tau |\lambda(\tau)|^2}{\int_0^\infty dz |\chi(z)|^2}  = N \,.
\ee
We can either think of $N$ as the norm of the sources with the normalization condition
\be
\label{norm1}
\int_0^\infty dz |\chi(z)|^2 = 1 \; ,
\ee
on the initial data, or we can think of $N$ as the inverse norm of the initial data if we set the $L_2$ norm of the source to be 1.\footnote{We could have alternatively considered the quantity $N'$ defined as the source norm with the normalization condition $\int_0^\infty dz \chi(z) = 1$ (as we would have for a delta function). We have checked that the qualitative behavior of $V_{min}(N')$ and $V_{min}(N)$ is similar.}

Defining a non-increasing function $V_{min}(N)$, as the minimum variance that we can obtain with norm less than or equal to $N$, we have three possibilities:
\begin{enumerate}
\item
The variance $V_{min}(N)$ goes to zero for some finite $N$. In this case, we can directly represent delta function initial data with sources having bounded $L^2$ norm.
\item
The variance $V_{min}(N)$ approaches 0 as we allow $N$ to $\infty$. In this case, by taking the amplitude of the initial data small enough, we can obtain initial data with arbitrarily small variance about $z_0$ using sources with $L_2$ norm smaller than some bound.
\item
The variance $V_{min}(N)$ approaches some nonzero constant as $N \to \infty$. In this case, we would conclude that even by taking the amplitude of the initial data small as the variance goes to zero, it is not possible to approximate a delta function with bounded sources.
\end{enumerate}
To determine $V_{min}(N)$ we can consider an action\footnote{We restrict to real sources for simplicity.}
\be
S = \int_0^\infty dz \chi^2(z) (z - z_0)^2 - \Lambda \int_0^\infty dz \chi^2(z)  - \beta \int d \omega \lambda^2(\omega) \; .
\ee
We can consider the extrema of this action as a function of $\Lambda$ and $\beta$. These extremize $\int_0^\infty dz \chi^2(z) (z - z_0)^2$ for fixed $\int_0^\infty dz \chi^2(z) $ and $\int d \omega \lambda^2(\omega)$; the function that minimizes $V$ for fixed $N$ will then correspond to some particular values of $\Lambda$ and $\beta$.

Varying the action with respect to $\lambda(\omega)$ and using \eqref{init3}, we get
\be
\label{vareq}
 \int_0^\infty dz \chi(z) ((z - z_0)^2 - \Lambda) e^{- \omega z} - \beta \lambda(\omega) =0
\ee
This gives
\be
\label{IFE}
\int_0^\infty d \rho (M_2(\omega, \rho) - \Lambda M_0(\omega, \rho)) \lambda(\rho) =  \beta \lambda(\omega)
\ee
where
\be
M_0(\omega,\rho) = \int dz e^{- \omega z} e^{-\rho z} = {1 \over \omega + \rho}
\ee
and
\bea
M_2(\omega,\rho) &=& \int dz e^{- \omega z} (z - z_0)^2 e^{-\rho z} \cr
&=& ({d \over d \rho} + z_0)^2  {1 \over \rho + \omega} \cr
&=& {2 \over (\omega + \rho)^3} - {2 z_0 \over (\omega + \rho)^2} + {z_0^2 \over \omega + \rho}
\eea
Solutions of equation (\ref{IFE}) are eigenfunctions $\lambda_{\Lambda,\beta}(\omega)$ corresponding to some allowed values of $\beta$. For each allowed $\beta$, we can compute
\bea
Z(\Lambda,\beta) &=& \int_0^\infty dz \chi_{\Lambda,\beta}^2(z) (z - z_0)^2 \cr
N_\chi(\Lambda,\beta) &=& \int_0^\infty dz \chi_{\Lambda,\beta}^2(z) \cr
N_\lambda(\Lambda, \beta) &=& \int d \omega \lambda_{\Lambda,\beta}^2(\omega)
\eea
From these, we have
\bea
V(\Lambda,\beta) &=& {Z(\Lambda,\beta) \over N_\chi(\Lambda,\beta)} = \Lambda + \beta {N_\lambda(\Lambda, \beta) \over N_\chi(\Lambda,\beta)} \cr
N(\Lambda, \beta) &=& {N_\lambda(\Lambda, \beta) \over N_\chi(\Lambda, \beta)} \; .
\eea
Finally, we have that $V_{min}(N)$ is the minimum of $V(\Lambda,\beta)$ over $\beta$ and $\Lambda$ subject to $N(\Lambda, \beta) = N$. We can think of this as the lower bound on the region of the $V-N$ plane covered by a parametric plot of $(V(\Lambda,\beta),N(\Lambda, \beta))$ as a function of $\Lambda$ and $\beta$.

It is straightforward to carry out this analysis numerically. We take $\lambda(\omega)$ to be piecewise constant on small intervals of size $\epsilon$ covering a range $[\omega_{min}, \omega_{max}]$. In this case, the integral equation (\ref{IFE}) becomes an ordinary matrix eigenvalue equation for a square matrix of size $\epsilon^{-1}\left(\omega_{max}-\omega_{min}\right)$.

\begin{figure}
\centering
\includegraphics[width= 0.7 \textwidth]{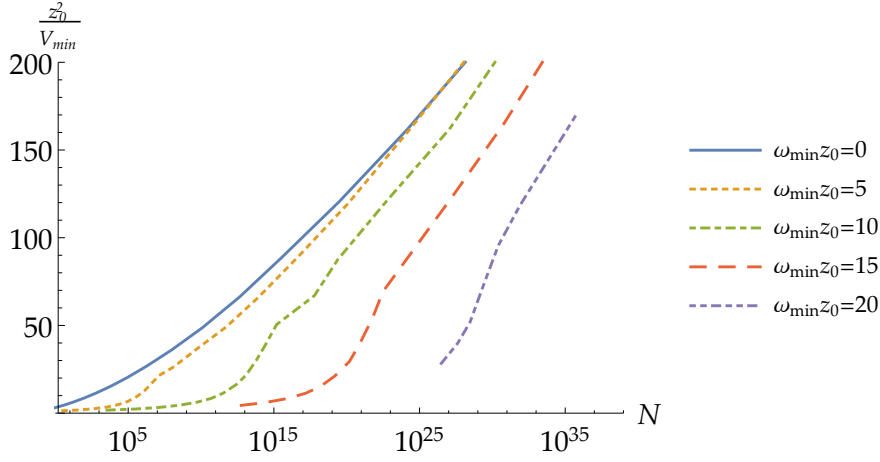}
\caption{Plot of $ z_0^2  \left(V_{min}\right)^{-1}$ vs $N$ on a semi-log scale, where $V_{min}$ is the minimum variance of an initial data function and $N$ is the ratio of the $L_2$ norm of sources to the $L_2$ norm of the initial data function. This figure uses $\omega_{max} z_0=40$ and $\epsilon z_0=\frac15$ and plots the results for values of $\omega_{min} z_0$ listed in the legend. The curve suggests that the minimum variance approaches zero as the norm of the sources is increased for any choice of $\omega_{min} z_0$.}
\label{NVfig}
\end{figure}

Our discretisation introduces dimensionless parameters $\epsilon z_0 \ll 1$, $\omega_{max} z_0 \gg 1$ and $\omega_{min} z_0 \ll 1$. In the next section we will also consider the case where $\omega_{min} z_0$ is held fixed.
Our explicit numerical results for $V_{min}(N)$ are plotted in figure \ref{NVfig} suggesting that $1/V_{min}$ is a function of $\log(N)$ that increases roughly linearly at large $\log(N)$.\footnote{Figure \ref{NVfig} uses $\epsilon z_0=\frac15$ and $\omega_{max} z_0 = 40$. Our numerical analysis found that the results in figure \ref{NVfig} were stable under decreasing $\epsilon z_0$ or increasing $\omega_{max} z_0$. The discretisation may seem fairly coarse, but the small $V_{min}$ regime is more sensitive to $\omega_{max} z_0$ and so a coarser discretisation allows us to reliably access lower variances while keeping the size of the required matrix under control.} For use in the next section, we note that qualitatively similar results are obtained if we keep $\omega_{min}$ fixed, i.e. we restrict the source to vanish for $\omega < \omega_{min}$.

These results are consistent with possibility 2 above: for any finite value of the $L_2$ norm for the sources, there is a minimum possible value for the variance and this minimum value appears to go to zero
as the norm is taken to infinity. This suggests that we can produce arbitrarily localized initial data, but we need to take the amplitude small as the variance becomes small if we want the $L_2$ norm of the sources to remain smaller than some particular value.

Recalling from formula (\ref{init3}) that the initial data function in this case is simply a Laplace transform of the Fourier transformed sources, our results here imply an interesting result for Laplace transforms: if $\chi : [0,\infty) \to \mathbb{R}$  is the Laplace transform of a function $\lambda$, then the variance of $\chi$ is bounded below by some decreasing positive function of the $L_2$ norm of $\lambda$. It may be interesting to understand this relationship more precisely.

\subsubsection{Fully localizing initial data}

Next, we extend the analysis of the previous section to the more general case where translation invariance in the spatial field theory directions is not assumed. Starting from the general expression
\be
\chi(z,x) = \int_{-\infty}^\infty {d \omega  d^{d-1} k \over (2 \pi)^{d-1}} e^{-\sqrt{\vec{k}^2 + \omega^2}z} e^{i \vec{k} \cdot \vec{x}} \lambda(\omega,\vec{k}) \; ,
\ee
we note first that the Fourier modes in the field theory directions of $\chi$ may be expressed as
\be
e^{i \vec{k} \cdot \vec{x}} F_k(z)
\ee
where $F_{\vec{k}}(z)$ is some linear combination of functions $e^{-\alpha z}$ with $\alpha \ge |k|$. That is, the functions $F_{\vec{k}}(z)$ are Laplace transforms of functions $\tilde{F}_{\vec{k}}(\omega)$ with the restriction that $\omega \ge |k|$.

Applying the same numerical methods as in the previous section, we have found that even with such a restriction on $\tilde{F}$
, it is still possible to produce a function $F_{\vec{k}}(z)$ with arbitrarily small variance so long as the sources are taken sufficiently large. Thus, for any $\vec{k}$, we can generate a function whose $x$-dependence is $e^{i \vec{k} \cdot \vec{x}}$ and whose variance in the $z$-direction is arbitrarily small. For our discussion below, let $N_{\omega_{min}}(V)$ be the minimum norm necessary to achieve a variance $V$ with a source whose support is in $[\omega_{min}, \infty)$.

We will now show that by taking linear combinations of mode functions constructed in this way, we can generate functions $\chi(\vec{x},z)$ with arbitrarily small variance by choosing sources $\lambda$ with sufficiently large (finite) norm. We will argue that this is possible to achieve using functions of the form $\chi(x,z) = X(x)Z(z)$.

Suppose we want to generate a function $X(x)Z(z)$ with variance less than $V_*$. First, we note that for large enough $k_0$, it is possible to find a function
\be
X(x) = \int_{|k| < k_0} d k A(k) e^{i \vec{k} \cdot \vec{x}}
\ee
written in terms of Fourier modes with $|k| < k_0$ such that $\int dx X(x)^2 = 1$ and $\int dx X(x)^2 x^2 < V_*/2$. Next, we can choose a function
\be
Z(z) = \int_{k_0}^\infty d\omega \lambda_{k_0}(\omega) e^{-\omega z} \,,
\ee
such that the variance is less than $V_*/2$ and source norm is less than or equal to $N_{k_0}(V_*/2)$, where $\lambda_{k_0}(\omega)$ is supported on $[k_0,\infty)$.

Finally, consider a source defined  as
\be
\lambda(k,\omega) = {A(k) \omega \over \sqrt{\omega^2 + k^2}} \lambda_{k_0} ( \sqrt{\omega^2 + k^2})  \; .
\ee
This gives
\bea
\chi(z,x) &=& \int dk d \omega {A(k) \omega \over \sqrt{\omega^2 + k^2} } \lambda_{k_0} (\sqrt{\omega^2 + k^2}) e^{i \vec{k} \cdot \vec{x}} e^{-\sqrt{k^2 + \omega^2} z} \cr
&=& \int_{|k|<k_0} dk  A(k) e^{i \vec{k} \cdot \vec{x}} \int_{k_0}^\infty d \alpha   \lambda_{k_0}(\alpha) e^{-\alpha z} \cr
&=& X(x) Z(z) \; .
\eea

The variance for such a function is
\bea
V &=& {\int dx dz \chi(x,z)^2 (x^2 + (z - z_0)^2) \over \int dx dz \chi(x,z)^2} \cr
&=& {{\int dx X(x)^2 x^2 \over \int dx X(x)^2} + {\int dz Z(z)^2 (z - z_0)^2 \over \int dz Z(z)^2}} \cr
&=& V_Z + V_X < V_*
\label{varsplit}
\eea
Finally, the norm of the sources is
\bea
N &=& \frac{\int dk d \omega \left[ \frac{A(k) \omega}{\sqrt{\omega^2 + k^2}}  \lambda_{k_0}( \sqrt{\omega^2 + k^2}) \right]^2 }{\int dx dz \chi(x,z)^2} \cr
&=& \frac{\int_{|k|<k_0} dk \int_{k_0}^\infty d \alpha {\sqrt{\alpha^2 - k^2} \over \alpha} A^2(k)\lambda_{k_0}^2(\alpha)}{\int dx dz \chi(x,z)^2} \cr
& \le & \left(\frac{\int_{|k|<k_0} dk A^2(k)}{\int dx X(x)^2} \right)
\left(\frac{\int_{k_0}^\infty d \alpha \lambda_{k_0}^2(\alpha)}{\int dz Z(z)^2} \right)\cr
& = &\frac{1}{2\pi} N_{k_0}(V_*/2) \; ,
\eea
where we have used Plancherel's theorem. Thus, we have seen that it is possible to obtain arbitrarily small variance for sufficiently large source norm, so long as this is possible in the translation-invariant case for sources with support in $[\omega_{min}, \infty)$.

The form of $N_{\omega_{min}}(V_*)$ is investigated in Figure \ref{NVfig}. We see that for a sufficiently small $\omega_{min}$, $N_{\omega_{min}}(V_*)$ asymptotes within the range of our numerics to the roughly linear relationship seen for $N_{0}(V_*)$ in the translation invariant analysis. The delay with which $N_{\omega_{min}}(V_*)$ reaches this asymptotic behaviour increases with $\omega_{min}$ and so our results are consistent with the conjecture that this is the generic asymptotic behaviour of $N_{\omega_{min}}(V_*)$ at smaller variances than our numerics can probe.
In any case, figure \ref{NVfig} provides evidence that small variances can be achieved at the cost of exponentially suppressing the amplitude independent of this observation.

\section{Absence of $\omega > |k|$ modes in CFT one-point functions.}\label{sec5}

An important feature of the perturbative results for the CFT one-point functions in terms of sources is that the one-point functions have no $|\omega| < |\vec{k}|$ modes. For holographic theories, this has been noted before in \cite{Bena:1999jv, Papadodimas:2012aq} at leading order in the $1/N$ expansion, and comes from the fact that these would correspond to perturbative bulk modes that diverge exponentially for large $z$.\footnote{This property leads to the result that CFT one-point functions integrated against certain non-trivial smearing functions will vanish for states in the ``code subspace''. This is related to the ``quantum error correction'' property for holographic states \cite{Almheiri:2014lwa, Mintun:2015qda, Freivogel:2016zsb}, i.e. that local bulk operators (which according to HKLL \cite{Hamilton:2006az} are equal to smeared local CFT operators at leading order) can be represented in multiple ways as boundary operators.} From a CFT perspective, the argument for why spacelike modes decouple in a large $N$ CFT is that they drop out of the two-point function (as we have already seen by explicit calculation), and at large $N$ all correlation functions factorize in terms of two-point functions. In this section, we show the absence of these $|\omega | < |\vec{k}|$ modes for one-point functions of CFT scalar quasi-primary operators quite generally, i.e., without assuming a large $N$ expansion.

We consider states
\be
\label{genstate}
|\Psi \rangle  = {\cal N} \sum_n c_n {\cal O}_n(\vec{x}_n, -\tau_n)|0 \rangle
\ee
obtained as linear combinations of states obtained by acting with scalar operators  ${\cal O}_n$ of some fixed scaling dimension $\Delta_n$ on the vacuum at spatial position $\vec{x}_n$ and Euclidean time $- \tau_n < 0$. A state
\be
{\cal O}_n(\vec{x}_n, -\tau_n)|0 \rangle
\ee
is the same as that obtained by inserting the operator ${\cal O}_n(\vec{x}_n, -\tau_n)$ into the Euclidean path-integral defining the vacuum state. We recall that any CFT state can be obtained by the insertion of some linear combination of operators of fixed scaling dimension at $\tau = - \infty$; here, by inserting scalar operators at arbitrary position, we can reproduce states obtained by inserting any scalar primary operator or any of its conformal descendants at $\tau = - \infty$.\footnote{To construct a completely general state, we must also include insertions primary operators in other $SO(d)$ representations and their descendants. We expect that it should be possible to generalize the arguments in this section to that general case.}

Now, for the general state (\ref{genstate}), consider the one-point function of a scalar operator ${\cal O}(x,t)$ of dimension $\Delta$ at Lorentzian time $t$. This is obtained by analytic continuation from the Euclidean three-point function. Keeping track only of the $x$ and $t$ dependence, we have
\be
\label{onepoint}
\langle \Psi | {\cal O}(x,t) | \Psi \rangle = \sum_{m,n} {A_{m,n} \over ((\vec{x} - \vec{x}_m)^2 + (\tau_m - it)^2)^{{1 \over 2}(\Delta + \Delta_m - \Delta_n)}((\vec{x} - \vec{x}_n)^2 + (\tau_n - it)^2)^{{1 \over 2}(\Delta + \Delta_n - \Delta_m)}} \; .
\ee
Using the identity
\be
{1 \over A^n B^m} = {\Gamma[m+n] \over \Gamma[m] \Gamma[n]} \int_0^1 ds {s^{m-1} (1-s)^{n-1} \over (sA + (1-s)B)^{m+n}} \; ,
\ee
we find that the $x$ and $t$ dependent function in (\ref{onepoint}) is a linear combination of terms of the form
\be
\label{genK}
{1 \over ((\vec{x} - \vec{x}_0)^2 + (\tau_0 - it)^2 + C^2)^\Delta}
\ee
where
\beas
\vec{x}_0 &=& s \vec{x}_m + (1-s) \vec{x}_n \cr
\tau_0 &=& s \tau_m - (1-s) \tau_n \cr
C^2 &=& s(1-s)((\vec{x}_m - \vec{x}_n)^2 + (\tau_m + \tau_n)^2)
\eeas
For $C=0$, the Fourier transform of (\ref{genK}) is the expression $K_\Delta(k,\omega,\tau)$ defined in (\ref{defKdelta}) and given explicitly in (\ref{Kdelta}). This contains no $|\omega| < |k|$ modes. If this remains true for $C \ne 0$, then we would conclude that one-point functions for general states have no $|\omega| < |k|$ modes. Naively, we could expand (\ref{genK}) in powers of $C$; the Fourier transformation of the $C^{2n}$ term is proportional to $K_{\Delta + n}$ and thus contains no  $|\omega| < |k|$ modes, but it's not clear that expanding, integrating, and resumming is justified here. In any case, the result of this procedure is
\be
K^C_\Delta(k,\omega,\tau) = 2 e^{- i \vec{k} \cdot \vec{x}_0} e^{- \tau \omega} \theta(\tau \omega) \theta(\omega^2 - k^2) {\pi^{{d \over 2} + 1} \over \Gamma[\Delta]} \left({ \omega^2 - k^2 \over 4 C^2} \right)^{{\Delta \over 2} - {d \over 4}}  J_{\Delta - {d \over 2}} \left[C\sqrt{\omega^2 - k^2}\right]
\ee

To check this, we evaluate the Fourier transformation directly using contour integration methods as in section 3. We have (for positive integer $\Delta$),
\bea
K^C_\Delta(k,\omega,\tau) &=& \int d^{d-1} \vec{x} dt e^{-i \vec{k} \cdot \vec{x}} e^{-i \omega t} {1 \over (C^2 + (\vec{x} - \vec{x}_0)^2+(\tau - it)^2)^\Delta} \\
&=& { \pi^{{d \over 2}-1} \Gamma[\Delta - {d \over 2} + 1] \over \Gamma[\Delta]} e^{- i \vec{k} \cdot \vec{x}_0} \int dt e^{-i \omega t} \int d x  {e^{-i |\vec{k}| x}  \over (C^2 + x^2+(\tau - it)^2)^{\Delta- {d \over 2} + 1}} \cr
&=& { \pi^{{d \over 2}-1} e^{- i \vec{k} \cdot \vec{x}_0} \over \Gamma[\Delta]}(-{d \over d C^2})^{\Delta- {d \over 2} }\int_{-\infty}^{\infty} dx e^{-i k x} \int_{-\infty}^{\infty} dt  e^{-i \omega t} {1 \over C^2 + x^2+(\tau - it)^2} \cr
&=&  { \pi^{{d \over 2}-1} e^{- \omega \tau} e^{- i \vec{k} \cdot \vec{x}_0} \over \Gamma[\Delta]}(-{d \over d C^2})^{\Delta- {d \over 2} }\int_{-\infty}^{\infty} dx e^{-i k x} \int_{-\infty+i\tau}^{\infty +i \tau} dt  e^{-i \omega t} {1 \over C^2 + x^2-t^2} \cr
&=&
{ 2 \pi^{d \over 2} e^{- \omega \tau} e^{- i \vec{k} \cdot \vec{x}_0} \theta(\omega \tau) \over \Gamma[\Delta]}(-{d \over d C^2})^{\Delta- {d \over 2} }\int_{-\infty}^{\infty} dx e^{-i k x}  {\sin(\omega\sqrt{C^2 + x^2}) \over \sqrt{C^2+ x^2}} \cr
&=&
{ 2 \pi^{{d \over 2}+1} e^{- \omega \tau} e^{- i \vec{k} \cdot \vec{x}_0} \theta(\omega \tau) \over \Gamma[\Delta]} \theta(\omega^2 - k^2) (-{d \over d C^2})^{\Delta- {d \over 2} } J_0(C\sqrt{\omega^2 - k^2})\cr
&=&
{ 2 \pi^{{d \over 2}+1} e^{- \omega \tau} e^{- i \vec{k} \cdot \vec{x}_0} \theta(\omega \tau) \over \Gamma[\Delta]} \theta(\omega^2 - k^2) \left({ \omega^2 - k^2 \over 4 C^2} \right)^{{\Delta \over 2} - {d \over 4}}  J_{\Delta - {d \over 2}} \left[C\sqrt{\omega^2 - k^2}\right] \nonumber
\eea
which reproduces the naive result above in any CFT, the one-point function of any scalar primary operator is built from Fourier modes with the restriction that $|\omega|^2 > k^2$. Roughly, this says that the spatial variation must not be greater than the variation in time. For example, we can't have a one-point function that is static and localized in space.

Our discussion above extends the observations in the existing literature \cite{Bena:1999jv, Hamilton:2006az, Papadodimas:2012aq} about the absence of spacelike modes from one-point functions to all orders in $1/N$, and indeed even to more general CFTs without a large $N$ expansion. However, as shown in \cite{Papadodimas:2012aq}, spacelike modes can be present in non-vanishing higher-point functions, for example in the thermal \emph{two-point function}. This is not in contradiction with the result discussed in this section, but demonstrates the need for care in interpreting our results.

\section{Discussion}\label{sec6}

In this paper, we have investigated in more detail the correspondence between states (\ref{PIstate}) defined by adding sources to the Euclidean path integral and the Lorentzian asymptotically AdS spacetimes that these states are dual to in holographic theories. We have presented evidence that at first order in perturbation theory, arbitrary perturbations to the background AdS spacetime can be well-represented by such path integral states. However, an interesting qualitative feature is that the validity of this perturbative approach depends both on the size of the bulk perturbations we wish to produce and on how localized these perturbations are.

An interesting generalization of this work would be to investigate states for which sources are added to the Euclidean path integral defining a mixed state -- for instance, the thermal state  -- instead of the vacuum path integral we considered here. States produced in this way should correspond to perturbations to the black hole or black brane geometries dual to the thermal state. Again, it would be interesting to understand the detailed map between sources and Lorentzian solutions, starting with the linearized analysis. Of course, one could also cast this problem in terms of the thermofield double, which points to the fact that there will be multiple different path-integral states which produce the same bulk initial data in the causal wedge of one side. Similarly, it might be interesting to consider the analogous calculation for subregions in the vacuum -- namely, if we were only to specify initial bulk data in the causal wedge of some (say, ball-shaped) CFT subregion (i.e. on a Cauchy surface in the correspondingly dual AdS hyperbolic black hole), then it is clear that several different path-integral states in the CFT would reproduce the required bulk data.

It would also be interesting to investigate which Lorentzian initial data can be generated via path integral states to higher orders in perturbation theory or non-perturbatively (e.g. via a numerical gravity calculation). For example, in a purely gravitational setting or for gravity coupled to a scalar field, it would be interesting to study numerically which spherically symmetric or translation-invariant initial data can be produced using sources with the same symmetry properties by solving the full nonlinear gravitational equations. In this case, we expect that certain sources will lead to singularities in the bulk Euclidean solution, so there will not be a one-to-one map between sources and smooth initial data. It is also not at all clear that we would be able to obtain arbitrary initial data by an appropriate choice of sources. Ultimately, we would like to understand whether any physical Lorentzian spacetime with a good classical description in a consistent quantum theory of gravity can be well-represented by states of the form (\ref{PIstate}) or if not, whether there is a nice characterization of which class of spacetimes these states correspond to.

\section*{Acknowledgements}
We would like to thank Tom Faulkner, Aitor Lewkowycz, and Mukund Rangamani for useful discussions.  The work of DM, AIR, and MVR  was supported in part by the Simons Foundations. The work of MVR and AIR was supported in part by the Natural Sciences and Engineering Research Council of Canada.

OP and CR wish to acknowledge support from the Simons Foundation (\#385592, Vijay Balasubramanian) through the It From Qubit Simons Collaboration. CR was also supported by the Belgian Federal Science Policy Office through the Interuniversity Attraction Pole P7/37, by FWO-Vlaanderen through projects G020714N and G044016N, and by Vrije Universiteit Brussel through the Strategic Research Program ``High-Energy Physics''.

\appendix
\section*{Appendix}
\section{Higher orders}\label{appA}

In section 3, we have obtained at first order in perturbation theory the relation between Euclidean sources, bulk initial data, and CFT one-point functions. In this appendix, we point out that there is a relatively simple formula valid at any order in perturbation theory for the sources required to produce any particular one-point function. This may be useful in extending the analysis of this paper to higher orders.

Our discussion makes use of the shadow operator formalism \cite{Ferrara:1972xe,Ferrara:1972ay,Ferrara:1972uq,Ferrara:1973vz,SimmonsDuffin:2012uy}.  We will work in Euclidean signature in this section and use the analytic continuations described above to extract the Lorentzian one-point functions.

We wish to find the sources required to set $\langle \Psi_\lambda | \cO_\alpha(x) | \Psi_\lambda \rangle = \rho_\alpha(x)$, where $\alpha$ labels primary operators.\footnote{Repeated labels should be summed over the primary operators in the CFT.} Expanding in the source,
\begin{align}
\label{eqn:1pt-def}
\rho_\alpha(x) &= \langle \Psi_\lambda | \cO_\alpha(x) | \Psi_\lambda \rangle
= \int [d \phi(\tau)] e^{-S_E - \int dy \lambda_\beta(y) \cO_\beta(y) }  \cO_\alpha(x) \\
&=   \int dy \lambda_{\beta}(y)  \langle  \cO_\alpha(x)  \cO_{\beta}(y) \rangle
+ \frac12\int dy_1 dy_2 \langle  \cO_\alpha(x)  \cO_{\beta_1}(y_1)  \cO_{\beta_2}(y_2)  \rangle  \lambda_{\beta_1}(y_1) \lambda_{\beta_2}(y_2) \\
&\quad+ \frac1{3!}\int dy_1 dy_2 dy_3 \langle  \cO_\alpha(x)  \cO_{\beta_1}(y_1)  \cO_{\beta_2}(y_2) \cO_{\beta_3}(y_3)  \rangle  \lambda_{\beta_1}(y_1) \lambda_{\beta_2}(y_2) \lambda_{\beta_3}(y_3)
+ O(\lambda^4),
\label{eqn:1pt-expansion}
\end{align}
where $\langle \ldots \rangle$ denotes the vacuum correlation function.

Similarly, the source can be expanded in terms of the target one-point function,
\begin{align}
\lambda_\beta(y) =  \int dx K_{\beta,\alpha}^{(1)}(y,x) \rho_\alpha(x) + \frac12 \int dx K_{\beta,\alpha_1,\alpha_2}^{(2)}(y,x_1,x_2) \rho_{\alpha_1}(x_1) \rho_{\alpha_2}(x_2)  .
+O(\rho^3)
\end{align}
Plugging into \eqref{eqn:1pt-expansion} determines the integral kernels $K^{(i)}$:
\begin{align*}
\rho_\alpha(x) &=  \int dy dx_1 \langle \cO_\alpha(x) \cO_{\beta}(y) \rangle K_{\beta,\alpha_1}^{(1)}(y,x_1) \rho_{\alpha_1}(x_1) \\\\
0 &= \frac12 \int dx_1 dx_2  \rho_{\alpha_1}(x_1) \rho_{\alpha_2}(x_2)
\Big[ \int dy \langle \cO_\alpha(x) \cO_{\beta}(y) \rangle K_{\beta,\alpha_1,\alpha_2}^{(2)}(y,x_1,x_2) \\
&\quad+\int dy_1 dy_2 \langle \cO_\alpha(x) \cO_{\beta_1}(y_1)  \cO_{\beta_2}(y_2)  \rangle  K_{\beta_1,\alpha_1}^{(1)}(y_1,x_1) K_{\beta_2,\alpha_2}^{(1)}(y_2,x_2)
\Big]  \\\\
0 &=   \frac{1}{3!} \int dx_1 dx_2  \rho_{\alpha_1}(x_1) \rho_{\alpha_2}(x_2) \rho_{\alpha_3}(x_3)
\Big[ \int dy \langle \cO_\alpha(x) \cO_{\beta}(y) \rangle K_{\beta,\alpha_1,\alpha_2,\alpha_3}^{(3)}(y,x_1,x_2,x_3) \\
&\quad+3\int dy_1 dy_2 \langle \cO_\alpha(x) \cO_{\beta_1}(y_1)  \cO_{\beta_2}(y_2)  \rangle  K_{\beta_1,\alpha_1}^{(1)}(y_1,x_1) K_{\beta_2,\alpha_2,\alpha_3}^{(2)}(y_2,x_2,x_3) \\
&\quad+  \int dy_1 dy_2 dy_3 \langle \cO_\alpha(x) \cO_{\beta_1}(y_1)  \cO_{\beta_2}(y_2) \cO_{\beta_3}(y_3)  \rangle  K_{\beta_1,\alpha_1}^{(1)}(y_1,x_1) K_{\beta_2,\alpha_2}^{(1)}(y_2,x_2) K_{\beta_3,\alpha_3}^{(1)}(y_3,x_3)
\Big]
\end{align*}

The first equation requires that $K_{\beta,\alpha}^{(1)}(y,x)$ be the inverse of the two-point function $\langle \cO_\alpha(x) \cO_{\beta}(y) \rangle$ as an integral kernel. Note that this two-point function is completely fixed by conformal symmetry and is diagonal in the $\alpha,\beta$ labels. Formal shadow operators can be defined by
\begin{align}
\tilde{\cO}_\beta (y)  = \int dx K_{\beta,\alpha}^{(1)}(y,x) \cO_\alpha(x),
\end{align}
which transform under conformal transformations as an operator of dimension $\tilde{\Delta}_\alpha = d- \Delta_\alpha$. This ensures that
\begin{align}
\langle \cO_\alpha(x) \tilde{\cO}_\beta(y) \rangle &= \delta^d (x-y) \\
\langle \tilde{\cO}_\alpha(x) \tilde{\cO}_\beta(y) \rangle &= K_{\beta,\alpha}^{(1)}(y,x).
\end{align}
This last equation tells us what $K^{(1)}$ is, since the two-point function is determined by conformal symmetry.\footnote{Equations in this section involving integral kernels or shadow operators should be understood to hold when integrated against suitable test functions.}
This definition is a bit different in spirit from that used in \cite{SimmonsDuffin:2012uy}, but can be checked to agree using results derived in Section 2 of that work.

Higher orders are straightforward:
\begin{align}
K_{\beta,\alpha_1,\alpha_2}^{(2)}(y,x_1,x_2)  &= - \langle \tilde\cO_\beta(y) \tilde\cO_{\alpha_1}(x_1)  \tilde\cO_{\alpha_2}(x_2)  \rangle \\
K_{\beta,\alpha_1,\alpha_2,\alpha_3}^{(3)}(y,x_1,x_2,x_3)  &= 3  \int dy_2 \langle \tilde\cO_\alpha(y) \tilde\cO_{\alpha_1}(x_1)  \cO_{\beta_2}(y_2)  \rangle
\langle \tilde\cO_{\beta_2}(y_2) \tilde\cO_{\alpha_2}(x_2)  \tilde\cO_{\alpha_3}(x_3)   \nonumber\\
&\qquad- \langle \tilde\cO_\beta(y) \tilde\cO_{\alpha_1}(x_1)  \tilde\cO_{\alpha_2}(x_2) \tilde\cO_{\alpha_3}(x_3)  \rangle.
\end{align}
$K^{(3)}$ involves a new ingredient, which is the main reason to introduce the shadow operator formalism. The OPE expansion allows us to fuse correlation functions of the form
\begin{align}
\int dx \langle \cO_1 \ldots \cO_m \cO_\alpha(x) \rangle \langle \tilde\cO_\alpha(x) \cO_{m+1} \ldots  \cO_n \rangle \Big|_P =  \langle \cO_1 \ldots \cO_n \rangle,
\end{align}
subject to a projection which eliminates the contributions from unphysical ``shadow conformal blocks''. This projection can be realised by picking out the part of the resulting integral with a particular monodromy behaviour for the positions of the operators $\cO_1 \ldots \cO_m$.
See \cite{SimmonsDuffin:2012uy} for a detailed discussion of how to do this projection. As noted there, this projection makes these expressions somewhat formal, but can be carried out in practice. Since this projection is trivial for $m=1$, all the integrals used in defining the source so far can be redefined to include such projections. Thus,
\begin{align}
K_{\beta,\alpha_1,\alpha_2,\alpha_3}^{(3)}(y,x_1,x_2,x_3)  &= 2 \langle \tilde\cO_\beta(y) \tilde\cO_{\alpha_1}(x_1)  \tilde\cO_{\alpha_2}(x_2) \tilde\cO_{\alpha_3}(x_3)  \rangle.
\end{align}

Using this approach, a formal expression for the source required to reproduce a given target one-point function to all orders is
\begin{align}
\lambda_\beta(y)= \left\langle \tilde\cO_\beta(y) \log \left(  1+\int dx \, \tilde\cO_\alpha(x) \rho_\alpha(x) \right) \right\rangle,
\end{align}
where this expression should be understood via its series expansion in powers of $\rho$.
It can be checked by plugging into \eqref{eqn:1pt-def},
\begin{align}
\langle \Psi_\lambda | \cO_\alpha(x) | \Psi_\lambda \rangle &= \left \langle   e^{\int dy \lambda_\beta(x) \cO_\beta(y)} \, \cO_\alpha(x) \right\rangle
= \left\langle
e^{\int dy \cO_\beta(y) \left\langle \tilde\cO_\beta(y) \log \left(  1+\int dx \, \tilde\cO_\alpha(x) \rho_\alpha(x) \right) \right\rangle }
\cO_\alpha(x) \right\rangle \\
=& \left\langle e^{ \log \left(  1+\int dx \, \tilde\cO_\alpha(x) \rho_\alpha(x) \right)} \cO_\alpha(x) \right\rangle
= \left\langle \left(  1+\int dx \, \tilde\cO_\alpha(x) \rho_\alpha(x) \right) \cO_\alpha(x) \right\rangle = \rho_\alpha(x).
\end{align}

\providecommand{\href}[2]{#2}\begingroup\raggedright\endgroup

\end{document}